\documentclass[%
 reprint,
 amsmath,amssymb,
 aps,
]{revtex4-1}

\usepackage{graphicx}% Include figure files
\usepackage{dcolumn}% Align table columns on decimal point
\usepackage{bm}% bold math
\usepackage{tabularx}
\usepackage{longtable}
\usepackage{threeparttable}
\usepackage{multirow}
\usepackage{rotating}
\usepackage[]{units}
\usepackage[T1]{fontenc}
\usepackage[version=3]{mhchem}
\usepackage{amsmath}
\usepackage{bm}% bold math
\usepackage{balance} 
\usepackage{float}
\usepackage{tablefootnote}
\usepackage{makecell}
\usepackage{flushend}
\newcommand{\Lower}[1]{\smash{\lower 1.5ex \hbox{#1}}}
\newcommand{\ab}{\bar{a}}
\newcommand{\ib}{\bar{i}}
\newcommand{\mc}{\multicolumn}

\begin{document}

\preprint{APS/AP1roG-noncovalent}

\title{Benchmarking the accuracy of seniority-zero wavefunction methods for non-covalent interactions}

\author{Filip Brz\k{e}k}
\affiliation
{Institute of Physics, Faculty of Physics, Astronomy and Informatics, Nicolaus Copernicus University in Torun, Grudziadzka 5, 87-100 Torun, Poland}
\author{Katharina Boguslawski}
\email{k.boguslawski@fizyka.umk.pl}
\affiliation
{Institute of Physics, Faculty of Physics, Astronomy and Informatics, Nicolaus Copernicus University in Torun, Grudziadzka 5, 87-100 Torun, Poland}
\affiliation
{Faculty of Chemistry, Nicolaus Copernicus University in Torun, Gagarina 7, 87-100 Torun, Poland}
\author{Pawe{\l} Tecmer}
\affiliation
{Institute of Physics, Faculty of Physics, Astronomy and Informatics, Nicolaus Copernicus University in Torun, Grudziadzka 5, 87-100 Torun, Poland}
\author{Piotr Szymon \.{Z}uchowski}
\email{pzuch@fizyka.umk.pl}
\affiliation
{Institute of Physics, Faculty of Physics, Astronomy and Informatics, Nicolaus Copernicus University in Torun, Grudziadzka 5, 87-100 Torun, Poland}

%\author{Pawe{\l} Tecmer}
%\affiliation{%
%Institute of Physics, Faculty of Physics, Astronomy and Informatics, Nicolaus Copernicus University in Torun, Grudziadzka 5, 87-100 Torun, Poland,
%\email{ptecmer@fizyka.umk.pl}
%}
%\author{Cristina E. Gonz\'alez-Espinoza}
%\affiliation{%
% Department of Chemistry and Chemical Biology, McMaster University, Hamilton, 1280 Main Street West, L8S 4M1, Canada}


%%%MAIN TEXT%%%%
\begin{abstract}
In this paper, we scrutinize the ability of seniority-zero wavefunction-based methods to model different types of non-covalent interactions, such as hydrogen bonds, dispersion, and mixed non-covalent interactions as well as prototypical model systems with various contributions of dynamic and static electron correlation effects. 
Specifically, we focus on the pair Coupled Cluster Doubles (pCCD) ansatz combined with two different flavours of dynamic energy corrections, (i) based on a perturbation theory correction and (ii) on a linearized coupled cluster ansatz on top of pCCD. 
We benchmark these approaches against the A24 data set [\v{R}ez{\'{a}}{\v{c}} and Hobza, J.~Chem.~Theory~Comput., 9, 2151-2155 (2013)] extrapolated to the basis set limit and some model non-covalent complexes that feature covalent bond breaking. 
By dissecting different types of interactions in the A24 data set within the Symmetry-Adapted Perturbation Theory (SAPT) framework, we demonstrate that pCCD can be classified as a dispersion-free method. 
Furthermore, we found that both flavours of post-pCCD approaches represent encouraging and computationally more efficient alternatives to standard electronic structure methods to model weakly-bound systems, resulting in small statistical errors. 
Finally, a linearized coupled cluster correction on top of pCCD proved to be most reliable for majority of investigated systems, featuring smaller non-parallelity errors compared to perturbation-theory-based approaches.
\end{abstract}
%%%%%%%%%%%%%%%%%%%%%%%%%%%%%%%%%%%%%%%%%%%%%%%%%%%%%%%%%%%%%%%%%%%%%%%%%%%%%%%%%%%%%%%%%%%%%%%%%%%%%%%%%%%%%%%%%%%%%%%%%%

\maketitle
%%%%%%%%%%%%%%%%%%%%%%%%%%%%%%%%%%%%%%%%%%%%%%%%%%%%%%%%%%%%%%%%%%%%%%%%%%%%%%%%%%%%%%%%%%%%%%%%%%%%%%%%%%%%%%%%%%%%%%%%%%
%%%%%%%%%%%%%%%%%%%%%%%%%%%%%%%%%%%%%%%%%%%%%%%%%%%%%%%%%%%%%%%%%%%%%%%%%%%%%%%%%%%%%%%%%%%%%%%%%%%%%%%%%%%%%%%%%%%%%%%%%%
\section{Introduction}
%%%%%%%%%%%%%%%%%%%%%%%%%%%%%%%%%%%%%%%%%%%%%%%%%%%%%%%%%%%%%%%%%%%%%%%%%%%%%%%%%%%%%%%%%%%%%%%%%%%%%%%%%%%%%%%%%%%%%%%%%%
%%%%%%%%%%%%%%%%%%%%%%%%%%%%%%%%%%%%%%%%%%%%%%%%%%%%%%%%%%%%%%%%%%%%%%%%%%%%%%%%%%%%%%%%%%%%%%%%%%%%%%%%%%%%%%%%%%%%%%%%%%
A reliable quantum chemical description of potential energy landscapes of chemical reactions is of crucial importance in chemistry, physics, biology, and astrophysics~\cite{mike-robb-review-1996,qc-in-biology,bowman-pes-2010,jankowski-science,szalay2012,pzuch-review-cold-chem-2018}. 
To predict the rates of chemical reactions, one has to know the dynamics of the nuclei, which is determined by the electronic energy at a given geometry. 
This involves, however, the ability to accurately calculate energy differences for molecular systems with stretched bonds and in multi-configurational states. 
Despite the significant progress in the development of multi-reference quantum chemistry methods in the past years~\cite{jeziorski-lmr-cc,adamowicz-mrcc,marti2010b,monika_mrcc,szalay2012,ors_ijqc,ors-lif-ttns,wouters-review,yanai-review,chanreview,evangelista-adaptive-ci}, it is still very challenging to tackle such problems. 
Conventional quantum chemistry methods that employ an active orbital space to describe static and non-dynamic correlation are difficult to use because of their complexity, non-blackbox character, and factorial scaling with respect to the system size. 
Hence, there is a great need for further development of electronic structure methods that allow us to accurately and reliably model bond-breaking situations without defining active spaces, but at the same time recover properly subtle effects such as dispersion interactions that originate from dynamic electron correlation.  

One promising alternative can be found in geminal-based methods combined with a proper dynamic energy correction~\cite{surjan1985,surjan-bond-2000,jeszenszki2014,gordon-vb-cc,frozen-pccd,kasia-lcc,reinhardt2017,ap1rog-ptx,pastorczak2017,hapka-jctc-2019}. 
Most commonly used is the Antisymmetric Product of Strongly orthogonal Geminals (APSG) model~\cite{hurley_1953,parr_1956,parks_1958,kutzelnigg_1964,kutzelnigg_1965,rassolov_2002}.  
One way to correct for the missing part of the dynamic correlation energy in APSG is to apply a multi-reference perturbation theory correction or a linearized coupled cluster ansatz developed by Surjan and coworkers~\cite{rosta2002,zoboki2013}.
A particularly interesting approach to model dynamic electron correlation effects in the APSG and Generalized Valence Bond (GVB~\cite{gvb1,gvb2}) models was developed by Pernal and coworkers~\cite{pernal-review-2018,pernal-apsg,chatterjee2012,pastorczak2015,chatterjee2016,pastorczak2017} and is based on the extended random-phase approximation. To this end, the fluctuation-dissipation theorem was used to connect the density fluctuations of electronic pairs with the intergeminal interaction energy. 
  
Another promising geminal-based method is the Antisymmetric Product of 1-reference orbital Geminal (AP1roG) model~\cite{limacher_2013,p-ccd}, also known as the pair Coupled Cluster Doubles (pCCD) ansatz~\cite{tamar-pcc,p-ccd}, which combined with an orbital optimization protocol~\cite{oo-ap1rog,ps2-ap1rog,ap1rog-jctc} represents a versatile tool to model strongly-correlated closed-shell systems at low computational cost.  
When the molecular orbitals are variationally optimized, the pCCD model becomes size-consistent and the resulting molecular basis is appropriate to model stretched bonds in contrast to standard canonical Hartree--Fock orbitals. 
As a consequence, potential-energy surfaces with a correct description of dissociation processes are obtained at mean-field-like cost.
Yet, the exponential form of the ansatz  allows for the desired (linear) scaling with the number of electrons (size-extensivity). 
This size-consistent and size-extensive methodology has proven to be useful in a wide scope of applications ranging from purely theoretical models~\cite{oo-ap1rog,piotrus_mol-phys,tamar-pcc,boguslawski2016} to real chemical problems including bond-breaking processes~\cite{pawel_jpca_2014,ps2-ap1rog,ap1rog-jctc,kasia_ijqc,ijqc-eratum,pawel_pccp2015,pccd-actinides,ola-book-chapter-2019}.
The missing fraction of dynamic correlation energy on top of pCCD can be added~\textit{a posteriori}, similar as for APSG, using density functional theory~\cite{garza2015,garza-pccp,garza_rpa}, many-body perturbation theory~\cite{piotrus_pt2,ap1rog-ptx}, and coupled cluster theory~\cite{kasia-lcc,frozen-pccd} corrections. 
Moreover, several extensions to pCCD have been proposed to model exited states~\cite{boguslawski2016targeting,boguslawski2016targeting-erratum,kasia-eom-pccd-lccsd}.  

Recently, some of us~\cite{ap1rog-ptx} developed a new family of dynamic energy corrections on top of a pCCD reference function, where the influence of a diagonal and off-diagonal zero-order Hamiltonian, a single- and multi-determinant wavefunction as dual, and different excitation operators used to construct the projection manifold were carefully examined~\cite{ap1rog-ptx}. 
Benchmarking those models against 15 reaction energies composed of weakly interacting closed-shell molecules shows that using an off-diagonal zero-order Hamiltonian and the full quantum-chemical Hamiltonian as perturbation operator reduces the error with respect to CR-CC(2,3) reference data to 2 kcal/mol. 
The application of a linearized coupled cluster singles and doubles correction on top of pCCD brings the results even closer to the reference values. 
Keeping in mind that these pCCD-based methods are also capable of modeling covalently stretched bonds in closed-shell systems, these findings motivate us to test their performance for more challenging weakly-interacting systems including covalent bond-breaking processes. 

Despite being much weaker in nature than forces bonding molecules internally (ionic and covalent), van der Waals or non-covalent interactions are responsible for a wide range of physical, chemical, and biological phenomena ranging from small molecules to complex systems in condensed phase~\cite{jeziorski1994,szalewicz2012}. 
Van der Waals forces might, for example, significantly influence molecular structures, phase transitions, and melting points. 
A reliable quantum mechanical description of small and medium-sized van der Waals complexes is nicely achieved within Symmetry-Adapted Perturbation Theory (SAPT)~\cite{sapt-review,jeziorski1994,moszynski1994,szalewicz2012}. 
Unfortunately, this beautiful methodology, as other methods designed specifically to model van der Waals complexes~\cite{grimme-van-der-waals-review}, has problems when static electron correlation effects have to be captured as, for instance, present in covalent bond breaking. 
%Standard quantum chemistry methods that are good for static electron correlation, such as, the multi-reference configuration interaction (MRCI~\cite{szalay2012}) and complete active space self-consistent field (CASSCF~\cite{werner_1985,roos_casscf,siegbahn_casscf}) theories, are on the other hand, not reliable for van der Waals interactions due to lack of size-consistency.
Standard quantum chemistry methods that are good for static electron correlation, such as, the multi-reference configuration interaction (MRCI~\cite{szalay2012}) and complete active space second-order perturbation theory (CASPT2~\cite{caspt21,caspt22}), are on the other hand, not reliable for van der Waals interactions due to lack of size-consistency.
Thus, we believe that our seniority-zero-based methods and their extensions represent a promising alternative that can bridge this gap.  

In this article, we will focus on non-covalent interactions and scrutinize (i) to what extend the pCCD method combined with an orbital optimization protocol can describe the interaction energy in weakly-bound systems and (ii) which of the dynamic energy corrections on top of pCCD is most suitable to account for the dispersion energy in these systems. 
Our starting point will be the A24 data set designed by \v{R}ez{\'{a}}{\v{c}} and Hobza~\cite{a24} to test different types of non-covalent interactions. 
This molecular test set is composed of 24 non-covalent complexes with all sorts of non-covalent interactions such as hydrogen bonds, mixed electrostatics/dispersion, and dispersion-dominated interactions including $\pi-\pi$ stacking. 
%Yet, the number of basis functions in these reference systems is small enough to be computed in house made programs as ours.
Furthermore, we investigate the performance of our newly developed methods for a few non-covalently interacting monomers such as the \ce{H_2}$\cdots$\ce{H_2}, \ce{H_2}$\cdots$LiH, and \ce{H_3^+}$\cdots$\ce{H_2} model complexes, in which the \ce{H-H}, \ce{Li-H}, and \ce{H-H} bonds are stretched, respectively.

This work is organized as follows. In section~\ref{sec:theory}, we briefly recapitulate our theoretical methods. Section~\ref{sec:comp} contains all computational details. Our numerical results are presented in section~\ref{sec:results}. Finally, in section~\ref{sec:conclusions} we deliver our conclusions and outlook.

%%%%%%%%%%%%%%%%%%%%%%%%%%%%%%%%%%%%%%%%%%%%%%%%%%%%%%%%%%%%%%%%%%%%%%%%%%%%%%%%%%%%%%%%%%%%%%%%%%%%%%%%%%%%%%%%%%%%%%%%%%
%%%%%%%%%%%%%%%%%%%%%%%%%%%%%%%%%%%%%%%%%%%%%%%%%%%%%%%%%%%%%%%%%%%%%%%%%%%%%%%%%%%%%%%%%%%%%%%%%%%%%%%%%%%%%%%%%%%%%%%%%%
\section{Theory}\label{sec:theory}
%%%%%%%%%%%%%%%%%%%%%%%%%%%%%%%%%%%%%%%%%%%%%%%%%%%%%%%%%%%%%%%%%%%%%%%%%%%%%%%%%%%%%%%%%%%%%%%%%%%%%%%%%%%%%%%%%%%%%%%%%%
%%%%%%%%%%%%%%%%%%%%%%%%%%%%%%%%%%%%%%%%%%%%%%%%%%%%%%%%%%%%%%%%%%%%%%%%%%%%%%%%%%%%%%%%%%%%%%%%%%%%%%%%%%%%%%%%%%%%%%%%%%
In this section we will briefly review the pCCD model~\cite{limacher_2013,tamar-pcc} combined with a variational orbital optimization procedure~\cite{oo-ap1rog,ap1rog-jctc} and post-pCCD dynamic energy corrections~\cite{piotrus_pt2,kasia-lcc,ap1rog-ptx} that have been investigated in this work.
For a complete picture of the methodology used, we refer to refs.~\citenum{limacher_2013,tamar-pcc,oo-ap1rog,ap1rog-jctc,piotrus_pt2,kasia-lcc,ap1rog-ptx}. 
Furthermore, we will scrutinize the definition of the interaction energy, its computational protocol as well as main issues concerning the accuracy of the obtained potential energy surfaces. 
%%%%%%%%%%%%%%%%%%%%%%%%%%%%%%%%%%%%%%%%%%%%%%%%%%%%%%%%%%%%%%%%%%%%%%%%%%%%%%%%%%%%%%%%%%%%%%%%%%%%%%%%%%%%%%%%%%%%%%%%%%
\subsection{pCCD and orbital optimization}
%%%%%%%%%%%%%%%%%%%%%%%%%%%%%%%%%%%%%%%%%%%%%%%%%%%%%%%%%%%%%%%%%%%%%%%%%%%%%%%%%%%%%%%%%%%%%%%%%%%%%%%%%%%%%%%%%%%%%%%%%%
In the pCCD ansatz, the cluster operator is restricted to electron pair excitations $T_{\rm p}$,
%%%%%%%%%%%%%%%%%%%%%%%%%%%%%%%%%%%%%%%%%%%%%%%%%%%%%%%%%%%%%%%%%%%%%%%%%%%%%%%%%%%%%%%%%%%%%%%%%%%%%%%%%%%%%%%%%%%%%%%%%%
%%%%%%%%%%%%%%%%%%%%%%%%%%%%%%%%%%%%%%%%%%%%%%%%%%%%%%%%%%%%%%%%%%%%%%%%%%%%%%%%%%%%%%%%%%%%%%%%%%%%%%%%%%%%%%%%%%%%%%%%%%
\begin{equation}\label{eq:ap1rog}
|{\rm pCCD}\rangle = \exp \left (  \sum_{i=1}^{\rm occ} \sum_{a=1}^{\rm virt} t_i^a a_a^{\dagger}  a_{\bar{a}}^{\dagger}a_{\bar{i}} a_{i}  \right )| 0 \rangle = e^{\hat{T}_{\rm p}} | 0 \rangle,
\end{equation}
%%%%%%%%%%%%%%%%%%%%%%%%%%%%%%%%%%%%%%%%%%%%%%%%%%%%%%%%%%%%%%%%%%%%%%%%%%%%%%%%%%%%%%%%%%%%%%%%%%%%%%%%%%%%%%%%%%%%%%%%%%
%%%%%%%%%%%%%%%%%%%%%%%%%%%%%%%%%%%%%%%%%%%%%%%%%%%%%%%%%%%%%%%%%%%%%%%%%%%%%%%%%%%%%%%%%%%%%%%%%%%%%%%%%%%%%%%%%%%%%%%%%%
where $a^\dagger_p$ and $a_p$ ($a^\dagger_{\bar{p}}$ and $a_{\bar{p}}$) are the electron creation and annihilation operators for $\alpha$ ($\beta$) electrons and $| 0 \rangle$ is some independent-particle wavefunction, for instance, the Hartree--Fock (HF) determinant.
In Eq.~\eqref{eq:ap1rog}, $\{t_i^a\}$ are the electron-pair amplitudes and $\hat{T}_{\rm p} = \sum_{i=1}^{\rm occ} \sum_{a=1}^{\rm virt} t_i^a a_a^{\dagger}  a_{\bar{a}}^{\dagger}a_{\bar{i}} a_{i} $ is the electron-pair excitation operator that excites an electron pair from an occupied $(i\ib)$ to a virtual orbital $(a\ab)$ with respect to $| 0 \rangle$.
Although being size-extensive by construction, pCCD is not size-consistent and thus does not yield reliable potential energy surfaces.
Size-consistency can be recovered if the one-particle basis functions are optimized.
This optimization can be done in a fully variational manner,~\cite{oo-ap1rog,ap1rog-jctc} analogous to orbital-optimized coupled cluster~\cite{scuseria1987} (OCC), or using approximate seniority-based projection techniques.~\cite{ps2-ap1rog,ap1rog-jctc}

In this work, the orbital basis has been optimized using a variational orbital optimization protocol.~\cite{oo-ap1rog,ap1rog-jctc}
For pCCD, the orbitals are then chosen to minimize the pCCD energy expression subject to the constraint that the pair amplitude equations are satisfied.
The energy Lagrangian reads~\cite{oo-ap1rog,ap1rog-jctc}
%%%%%%%%%%%%%%%%%%%%%%%%%%%%%%%%%%%%%%%%%%%%%%%%%%%%%%%%%%%%%%%%%%%%%%%%%%%%%%%%%%%%%%%%%%%%%%%%%%%%%%%%%%%%%%%%%%%%%%%%%%
%%%%%%%%%%%%%%%%%%%%%%%%%%%%%%%%%%%%%%%%%%%%%%%%%%%%%%%%%%%%%%%%%%%%%%%%%%%%%%%%%%%%%%%%%%%%%%%%%%%%%%%%%%%%%%%%%%%%%%%%%%
\begin{equation}\label{eq:voo}
\mathcal{L} = \langle 0 |  e^{-\hat{T}_{\rm p}}e^{\bm \kappa} \hat{H} e^{- \bm \kappa}  e^{\hat{T}_{\rm p}} | 0 \rangle +
              \sum_{i,a} \lambda_i^a \langle \Phi_{i \bar{i}}^{a \bar{a}} | e^{-\hat{T}_{\rm p}}e^{\bm \kappa} \hat{H} e^{- \bm \kappa}  e^{\hat{T}_{\rm p}} | 0 \rangle,
\end{equation}
%%%%%%%%%%%%%%%%%%%%%%%%%%%%%%%%%%%%%%%%%%%%%%%%%%%%%%%%%%%%%%%%%%%%%%%%%%%%%%%%%%%%%%%%%%%%%%%%%%%%%%%%%%%%%%%%%%%%%%%%%%
%%%%%%%%%%%%%%%%%%%%%%%%%%%%%%%%%%%%%%%%%%%%%%%%%%%%%%%%%%%%%%%%%%%%%%%%%%%%%%%%%%%%%%%%%%%%%%%%%%%%%%%%%%%%%%%%%%%%%%%%%%
where $\{\lambda_i^a\}$ are the Lagrange multipliers.
In the above equation, $\bm \kappa$ is the generator of orbital rotations,
%%%%%%%%%%%%%%%%%%%%%%%%%%%%%%%%%%%%%%%%%%%%%%%%%%%%%%%%%%%%%%%%%%%%%%%%%%%%%%%%%%%%%%%%%%%%%%%%%%%%%%%%%%%%%%%%%%%%%%%%%%
%%%%%%%%%%%%%%%%%%%%%%%%%%%%%%%%%%%%%%%%%%%%%%%%%%%%%%%%%%%%%%%%%%%%%%%%%%%%%%%%%%%%%%%%%%%%%%%%%%%%%%%%%%%%%%%%%%%%%%%%%%
\begin{equation}\label{eq:kappa}
\bm{\kappa} = \sum_{p>q} \kappa_{pq} (a^\dagger_p a_q - a^\dagger_q a_p),
\end{equation}
%%%%%%%%%%%%%%%%%%%%%%%%%%%%%%%%%%%%%%%%%%%%%%%%%%%%%%%%%%%%%%%%%%%%%%%%%%%%%%%%%%%%%%%%%%%%%%%%%%%%%%%%%%%%%%%%%%%%%%%%%%
%%%%%%%%%%%%%%%%%%%%%%%%%%%%%%%%%%%%%%%%%%%%%%%%%%%%%%%%%%%%%%%%%%%%%%%%%%%%%%%%%%%%%%%%%%%%%%%%%%%%%%%%%%%%%%%%%%%%%%%%%%
where $(\kappa_{pq})$ is a skew-symmetric matrix and transforms into a new orthogonal basis with a transformation $U = e^{-\bm \kappa}$ and $e^{\bm \kappa}\hat{H}e^{-\bm \kappa}$ is the Hamiltonian in the rotated basis.
We should note that the above sum runs over all orbital indices. Thus, in orbital-optimized pCCD, the occupied--occupied and virtual--virtual orbital rotations are non-redundant and have to be considered in the orbital optimization scheme.
In conventional CC theory, only the occupied--virtual rotations are non-redundant~\cite{scuseria1987}.

The requirement that the partial derivative of $\mathcal{L}$ with respect to the Lagrange multipliers $\{\lambda_i^a\}$ evaluated for the current orbitals is stationary results in the standard set of equations for the pCCD amplitudes $\{t_i^a\}$, 
%%%%%%%%%%%%%%%%%%%%%%%%%%%%%%%%%%%%%%%%%%%%%%%%%%%%%%%%%%%%%%%%%%%%%%%%%%%%%%%%%%%%%%%%%%%%%%%%%%%%%%%%%%%%%%%%%%%%%%%%%%
%%%%%%%%%%%%%%%%%%%%%%%%%%%%%%%%%%%%%%%%%%%%%%%%%%%%%%%%%%%%%%%%%%%%%%%%%%%%%%%%%%%%%%%%%%%%%%%%%%%%%%%%%%%%%%%%%%%%%%%%%%
\begin{equation}\label{eq:ap1rog-c}
\frac{\partial \mathcal{L}}{\partial \lambda_i^a}\Big\vert_{\bm \kappa=0} =  \langle \Phi_{i \bar{i}}^{a \bar{a}} |e^{-\hat{T}_{\rm p}}\hat{H} e^{\hat{T}_{\rm p}} | 0 \rangle = 0,
\end{equation}
%%%%%%%%%%%%%%%%%%%%%%%%%%%%%%%%%%%%%%%%%%%%%%%%%%%%%%%%%%%%%%%%%%%%%%%%%%%%%%%%%%%%%%%%%%%%%%%%%%%%%%%%%%%%%%%%%%%%%%%%%%
%%%%%%%%%%%%%%%%%%%%%%%%%%%%%%%%%%%%%%%%%%%%%%%%%%%%%%%%%%%%%%%%%%%%%%%%%%%%%%%%%%%%%%%%%%%%%%%%%%%%%%%%%%%%%%%%%%%%%%%%%%
while the stationary requirement of $\mathcal{L}$ with respect to the geminal coefficients, $\frac{\partial \mathcal{L}}{\partial {t_i^a}}\vert_{\bm \kappa=0} = 0$, leads to a set of equations for the Lagrange multipliers, analogous to the $\Lambda$-equations in CC theory,
%%%%%%%%%%%%%%%%%%%%%%%%%%%%%%%%%%%%%%%%%%%%%%%%%%%%%%%%%%%%%%%%%%%%%%%%%%%%%%%%%%%%%%%%%%%%%%%%%%%%%%%%%%%%%%%%%%%%%%%%%%
%%%%%%%%%%%%%%%%%%%%%%%%%%%%%%%%%%%%%%%%%%%%%%%%%%%%%%%%%%%%%%%%%%%%%%%%%%%%%%%%%%%%%%%%%%%%%%%%%%%%%%%%%%%%%%%%%%%%%%%%%%
\begin{align}\label{eq:ap1rog-l}
\frac{\partial \mathcal{L}}{\partial t_i^a}\Big\vert_{\bm \kappa=0} =\, &\langle 0| e^{-\hat{T}_{\rm p}} [ \hat{H}, a^\dagger_a a^\dagger_{\bar{a}} a_{\bar{i}} a_i ] e^{\hat{T}_{\rm p}} | 0 \rangle \nonumber \\
                                              &+\sum_{jb} \lambda_j^b \langle \Phi_{j \bar{j}}^{b \bar{b}} |  e^{-\hat{T}_{\rm p}} [ \hat{H}, a^\dagger_a a^\dagger_{\bar{a}} a_{\bar{i}} a_i  ] e^{\hat{T}_{\rm p}}| 0 \rangle.
\end{align}
%%%%%%%%%%%%%%%%%%%%%%%%%%%%%%%%%%%%%%%%%%%%%%%%%%%%%%%%%%%%%%%%%%%%%%%%%%%%%%%%%%%%%%%%%%%%%%%%%%%%%%%%%%%%%%%%%%%%%%%%%%
%%%%%%%%%%%%%%%%%%%%%%%%%%%%%%%%%%%%%%%%%%%%%%%%%%%%%%%%%%%%%%%%%%%%%%%%%%%%%%%%%%%%%%%%%%%%%%%%%%%%%%%%%%%%%%%%%%%%%%%%%%
The variational orbital gradient is the partial derivative of $\mathcal{L}$ with respect to the orbital rotation coefficients $\{\kappa_{pq}\}$,
%%%%%%%%%%%%%%%%%%%%%%%%%%%%%%%%%%%%%%%%%%%%%%%%%%%%%%%%%%%%%%%%%%%%%%%%%%%%%%%%%%%%%%%%%%%%%%%%%%%%%%%%%%%%%%%%%%%%%%%%%%
%%%%%%%%%%%%%%%%%%%%%%%%%%%%%%%%%%%%%%%%%%%%%%%%%%%%%%%%%%%%%%%%%%%%%%%%%%%%%%%%%%%%%%%%%%%%%%%%%%%%%%%%%%%%%%%%%%%%%%%%%%
\begin{align}\label{eq:voo-grad}
\frac{\partial \mathcal{L}}{\partial \kappa_{pq}}\Big\vert_{\bm \kappa=0} = g_{pq} = &\langle 0 | e^{-\hat{T}_{\rm p}} [(a^\dagger_p a_q - a^\dagger_q a_p),\hat{H}]e^{\hat{T}_{\rm p}} | 0 \rangle  \nonumber \\
               &+ \sum_{i,a} \lambda_i^a \big( \langle \Phi_{i \bar{i}}^{a \bar{a}} | e^{-\hat{T}_{\rm p}} [(a^\dagger_p a_q - a^\dagger_q a_p), \hat{H} ]e^{\hat{T}_{\rm p}} | 0\rangle \big).
\end{align}
%%%%%%%%%%%%%%%%%%%%%%%%%%%%%%%%%%%%%%%%%%%%%%%%%%%%%%%%%%%%%%%%%%%%%%%%%%%%%%%%%%%%%%%%%%%%%%%%%%%%%%%%%%%%%%%%%%%%%%%%%%
%%%%%%%%%%%%%%%%%%%%%%%%%%%%%%%%%%%%%%%%%%%%%%%%%%%%%%%%%%%%%%%%%%%%%%%%%%%%%%%%%%%%%%%%%%%%%%%%%%%%%%%%%%%%%%%%%%%%%%%%%%
To obtain the unitary transformation matrix $U$, we expand the energy Lagrangian as a function of $\bm \kappa$ up to second order
%%%%%%%%%%%%%%%%%%%%%%%%%%%%%%%%%%%%%%%%%%%%%%%%%%%%%%%%%%%%%%%%%%%%%%%%%%%%%%%%%%%%%%%%%%%%%%%%%%%%%%%%%%%%%%%%%%%%%%%%%%
%%%%%%%%%%%%%%%%%%%%%%%%%%%%%%%%%%%%%%%%%%%%%%%%%%%%%%%%%%%%%%%%%%%%%%%%%%%%%%%%%%%%%%%%%%%%%%%%%%%%%%%%%%%%%%%%%%%%%%%%%%
\begin{equation}
\mathcal{L}^{(2)}({\bm \kappa}) = \mathcal{L}^{(0)} +  {\bm \kappa} ^\dagger {\bm g} + \frac{1}{2}  {\bm \kappa} ^\dagger {\bm A}  {\bm \kappa},
\end{equation}
%%%%%%%%%%%%%%%%%%%%%%%%%%%%%%%%%%%%%%%%%%%%%%%%%%%%%%%%%%%%%%%%%%%%%%%%%%%%%%%%%%%%%%%%%%%%%%%%%%%%%%%%%%%%%%%%%%%%%%%%%%
%%%%%%%%%%%%%%%%%%%%%%%%%%%%%%%%%%%%%%%%%%%%%%%%%%%%%%%%%%%%%%%%%%%%%%%%%%%%%%%%%%%%%%%%%%%%%%%%%%%%%%%%%%%%%%%%%%%%%%%%%%
where $\bm A$ is the molecular orbital Hessian. Thus, minimizing the Lagrangian with respect to $\{\kappa_{pq}\}$ leads to the well-known equation for the orbital rotation coefficients
%%%%%%%%%%%%%%%%%%%%%%%%%%%%%%%%%%%%%%%%%%%%%%%%%%%%%%%%%%%%%%%%%%%%%%%%%%%%%%%%%%%%%%%%%%%%%%%%%%%%%%%%%%%%%%%%%%%%%%%%%%
%%%%%%%%%%%%%%%%%%%%%%%%%%%%%%%%%%%%%%%%%%%%%%%%%%%%%%%%%%%%%%%%%%%%%%%%%%%%%%%%%%%%%%%%%%%%%%%%%%%%%%%%%%%%%%%%%%%%%%%%%%
\begin{equation}\label{eq:kappa}
{\bm \kappa} = -{\bm A}  {\bm g}.
\end{equation}
%%%%%%%%%%%%%%%%%%%%%%%%%%%%%%%%%%%%%%%%%%%%%%%%%%%%%%%%%%%%%%%%%%%%%%%%%%%%%%%%%%%%%%%%%%%%%%%%%%%%%%%%%%%%%%%%%%%%%%%%%%
%%%%%%%%%%%%%%%%%%%%%%%%%%%%%%%%%%%%%%%%%%%%%%%%%%%%%%%%%%%%%%%%%%%%%%%%%%%%%%%%%%%%%%%%%%%%%%%%%%%%%%%%%%%%%%%%%%%%%%%%%%
After the orbital gradient and Hessian are evaluated (we employed a diagonal approximation of the exact orbital Hessian), the matrix representation of $\kappa$ can be determined from the above equation.
The transformation matrix is then approximated to second order as $U \approx 1 - \bm{\kappa} + \frac{1}{2} \bm{\kappa}^\dagger \bm{\kappa}$ and orthogonalized.
This transformation matrix is used to transform the current orbitals into the new basis.
The above steps (evaluation of eqs.~\eqref{eq:ap1rog-c}, \eqref{eq:ap1rog-l}, \eqref{eq:voo-grad}, and \eqref{eq:kappa}) are repeated until convergence, for instance, in the energy and orbital gradient, is reached. No convergence acceleration techniques, such as the Direct Inversion of the Iterative Subspace, were employed.

%%%%%%%%%%%%%%%%%%%%%%%%%%%%%%%%%%%%%%%%%%%%%%%%%%%%%%%%%%%%%%%%%%%%%%%%%%%%%%%%%%%%%%%%%%%%%%%%%%%%%%%%%%%%%%%%%%%%%%%%%%
\subsection{Dynamic correlation energy corrections on top of pCCD}
%%%%%%%%%%%%%%%%%%%%%%%%%%%%%%%%%%%%%%%%%%%%%%%%%%%%%%%%%%%%%%%%%%%%%%%%%%%%%%%%%%%%%%%%%%%%%%%%%%%%%%%%%%%%%%%%%%%%%%%%%%
The pCCD wavefunction allows us to reliably model non-dynamic/static electron correlation effects.~\cite{}
It misses, however, a large fraction of the dynamic correlation energy, which can be attributed to excitations beyond electron pairs.
In the following, we will briefly discuss corrections on top of pCCD that allow us to include dynamic correlation \textit{a posteriori}, focusing on different flavours of perturbation theory models and a linearized coupled cluster correction. 
An in depth discussion of all investigated corrections can be found in refs.~\citenum{piotrus_pt2,kasia-lcc,ap1rog-ptx}.
%%%%%%%%%%%%%%%%%%%%%%%%%%%%%%%%%%%%%%%%%%%%%%%%%%%%%%%%%%%%%%%%%%%%%%%%%%%%%%%%%%%%%%%%%%%%%%%%%%%%%%%%%%%%%%%%%%%%%%%%%%
\subsubsection{Perturbation theory models with a pCCD reference function}
%%%%%%%%%%%%%%%%%%%%%%%%%%%%%%%%%%%%%%%%%%%%%%%%%%%%%%%%%%%%%%%%%%%%%%%%%%%%%%%%%%%%%%%%%%%%%%%%%%%%%%%%%%%%%%%%%%%%%%%%%%
The perturbation theory models that have been developed for a pCCD reference function use Rayleigh--Schr\"odinger perturbation theory (RSPT) of second order.
In all approaches, the $\hat{H}_0$ Hamiltonian is chosen to be the inactive Fock matrix (with and without off-diagonal elements),
%%%%%%%%%%%%%%%%%%%%%%%%%%%%%%%%%%%%%%%%%%%%%%%%%%%%%%%%%%%%%%%%%%%%%%%%%%%%%%%%%%%%%%%%%%%%%%%%%%%%%%%%%%%%%%%%%%%%%%%%%%
\begin{align}\label{eq:fock}
    (\hat{H}_0)_N  =  \hat{F}_N &\nonumber  \\ 
    = \sum_{pq} \left(h_{pq} + \sum_i^{\rm occ} ( \langle pi||qi \rangle + \langle pi | qi\rangle)\right) \{a^\dagger_p a_q\} &\nonumber \\  = \sum_{pq} f_{pq} \{a^\dagger_p a_q\},
\end{align}
%%%%%%%%%%%%%%%%%%%%%%%%%%%%%%%%%%%%%%%%%%%%%%%%%%%%%%%%%%%%%%%%%%%%%%%%%%%%%%%%%%%%%%%%%%%%%%%%%%%%%%%%%%%%%%%%%%%%%%%%%%
where $\langle pi||qi \rangle $ are the two-electron integrals in physicists'  notation containing the Coulomb $\langle pi|qi \rangle $ and exchange $\langle pi|iq \rangle $ terms.
Note that we have taken the zero-order Hamiltonian in its normal-product form (see also Ref.~\citenum{ap1rog-ptx} for more details).
Based on the choice of $ (\hat{H}_0)_N $ (with or without off-diagonal elements), the perturbation is adjusted accordingly,
%%%%%%%%%%%%%%%%%%%%%%%%%%%%%%%%%%%%%%%%%%%%%%%%%%%%%%%%%%%%%%%%%%%%%%%%%%%%%%%%%%%%%%%%%%%%%%%%%%%%%%%%%%%%%%%%%%%%%%%%%%
\begin{equation}\label{eq:vn}
\hat{V}_N^\prime = \hat{H}_N-(\hat{H}_0)_N - E_{\mathrm{corr}}^{(0)} = \hat{V}_N - E_{\mathrm{corr}}^{(0)}.
\end{equation}
%%%%%%%%%%%%%%%%%%%%%%%%%%%%%%%%%%%%%%%%%%%%%%%%%%%%%%%%%%%%%%%%%%%%%%%%%%%%%%%%%%%%%%%%%%%%%%%%%%%%%%%%%%%%%%%%%%%%%%%%%%
In the above equation, the perturbation operator has been shifted by the pCCD correlation energy (indicated by the $\prime$) so that the first order correction to the energy vanishes,
%%%%%%%%%%%%%%%%%%%%%%%%%%%%%%%%%%%%%%%%%%%%%%%%%%%%%%%%%%%%%%%%%%%%%%%%%%%%%%%%%%%%%%%%%%%%%%%%%%%%%%%%%%%%%%%%%%%%%%%%%%
\begin{equation}\label{eq:e01}
            E^{(0)}+E^{(1)} = \frac{1}{\langle \tilde{\Psi}\vert \mathrm{pCCD}\rangle} \langle \tilde{\Psi} \vert \hat{V}_N^\prime \vert \mathrm{pCCD} \rangle = 0,
\end{equation}
%%%%%%%%%%%%%%%%%%%%%%%%%%%%%%%%%%%%%%%%%%%%%%%%%%%%%%%%%%%%%%%%%%%%%%%%%%%%%%%%%%%%%%%%%%%%%%%%%%%%%%%%%%%%%%%%%%%%%%%%%%
where $\langle \tilde{\Psi}\vert$ is the dual state.
The second-order energy can be calculated from the well-known expression of RSPT containing the first-order correction to the wavefunction $| \Psi^{(1)}\rangle$ and the shifted normal-product perturbation Hamiltonian,
%%%%%%%%%%%%%%%%%%%%%%%%%%%%%%%%%%%%%%%%%%%%%%%%%%%%%%%%%%%%%%%%%%%%%%%%%%%%%%%%%%%%%%%%%%%%%%%%%%%%%%%%%%%%%%%%%%%%%%%%%%
\begin{equation}\label{eq:e2}
            E^{(2)} = \frac{ \langle \tilde{\Psi} \vert \hat{V}_N^\prime \vert \Psi^{(1)} \rangle }{\langle \tilde{\Psi}\vert \mathrm{pCCD}\rangle}.
\end{equation}
%%%%%%%%%%%%%%%%%%%%%%%%%%%%%%%%%%%%%%%%%%%%%%%%%%%%%%%%%%%%%%%%%%%%%%%%%%%%%%%%%%%%%%%%%%%%%%%%%%%%%%%%%%%%%%%%%%%%%%%%%%
In all PT2 models presented in Ref.~\citenum{ap1rog-ptx}, the first-order correction to the wavefunction is written as a linear expansion of Slater determinants,
%%%%%%%%%%%%%%%%%%%%%%%%%%%%%%%%%%%%%%%%%%%%%%%%%%%%%%%%%%%%%%%%%%%%%%%%%%%%%%%%%%%%%%%%%%%%%%%%%%%%%%%%%%%%%%%%%%%%%%%%%%
        \begin{equation}\label{eq:psi1}
            \vert \Psi^{(1)} \rangle = \hat{T} \vert 0 \rangle,
        \end{equation}
%%%%%%%%%%%%%%%%%%%%%%%%%%%%%%%%%%%%%%%%%%%%%%%%%%%%%%%%%%%%%%%%%%%%%%%%%%%%%%%%%%%%%%%%%%%%%%%%%%%%%%%%%%%%%%%%%%%%%%%%%%
where $\hat{T}$ is some excitation operator that creates states that are orthogonal to the pCCD reference wavefunction, that is, we have $\langle {\Psi^{(1)}} \vert \mathrm{pCCD} \rangle = 0$.
Specifically, $\hat{T}$ is restricted to double excitations without pair excitations $\hat{T}_2^\prime$ (indicated by the $\prime$) as well as singles and doubles excitations (again without electron-pair excitations) $\hat{T}_1+\hat{T}_2^\prime$.
The PT2 amplitudes are determined by solving a set of equations,
%%%%%%%%%%%%%%%%%%%%%%%%%%%%%%%%%%%%%%%%%%%%%%%%%%%%%%%%%%%%%%%%%%%%%%%%%%%%%%%%%%%%%%%%%%%%%%%%%%%%%%%%%%%%%%%%%%%%%%%%%%
        \begin{equation}\label{eq:pt}
            \sum_p t_p \langle \overline{\Phi_q} \vert (\hat{H}_0)_N \vert \Phi_p \rangle + \langle \overline{\Phi_q} \vert \hat{V}_N^\prime \vert \mathrm{pCCD} \rangle = 0,
        \end{equation}
%%%%%%%%%%%%%%%%%%%%%%%%%%%%%%%%%%%%%%%%%%%%%%%%%%%%%%%%%%%%%%%%%%%%%%%%%%%%%%%%%%%%%%%%%%%%%%%%%%%%%%%%%%%%%%%%%%%%%%%%%%
where the sum runs over all Slater determinants included in the expansion Eq.~\eqref{eq:psi1} and the bar over the bra-states indicates that the final equations are spin-summed.

Based on the choice of the zero-order Hamiltonian, the excitation operator $\hat{T}$, and the dual state $\langle \tilde{\Psi} \vert $, we can derive different PT2 corrections on top of a pCCD reference function.
Various, recently presented PT2 models~\cite{piotrus_pt2,ap1rog-ptx} are summarized in Table~\ref{tab:pt} and scrutinized in Ref.~\citenum{ap1rog-ptx}.
Specifically, the nature of the dual state is abbreviated by \textit{SD} (single determinant) or \textit{MD} (multi determinants), while the shape of the zero-order Hamiltonian is indicated by a little letter (d: diagonal $\hat{F}_N$; o: off-diagonal $\hat{F}_N$).
The choice of the excitation operator is given in parentheses after each acronym and covers double (d) and single and double (sd) excitations.
We should note that in the original PT2b model,~\cite{piotrus_pt2} pair excitations are included in the first-order correction to the wavefunction Eq.~\eqref{eq:psi1} and hence $\hat{T}_2$ includes all double excitations.
To distinguish between $\hat{T}_2$ and $\hat{T}_2^\prime$ in the PT2b method, the exclusion of pair excitations is explicitly mentioned in parentheses, \textit{e.g.}, (d\textbackslash p) indicates that $\hat{T}_2^\prime$ has been chosen as excitation operator.
%%%%%%%%%%%%%%%%%%%%%%%%%%%%%%%%%%%%%%%%%%%%%%%%%%%%%%%%%%%%%%%%%%%%%%%%%%%%%%%%%%%%%%%%%%%%%%%%%%%%%%%%%%%%%%%%%%%%%%%%%%
%  Table 1
%%%%%%%%%%%%%%%%%%%%%%%%%%%%%%%%%%%%%%%%%%%%%%%%%%%%%%%%%%%%%%%%%%%%%%%%%%%%%%%%%%%%%%%%%%%%%%%%%%%%%%%%%%%%%%%%%%%%%%%%%%
\begin{table}
\begin{center}
%\begin{table}[htdp]
\caption{Summary of all PT models with a pCCD reference function investigated in this work. $\hat{H}_0$: zero-order Hamiltonian. $\hat{V}$: perturbation. $\langle \tilde{\Psi} \vert$: dual state. $\hat{T}$: excitation operator. $\hat{F}_N^{\mathrm{d}}$: diagonal part of $\hat{F}_N$. $\hat{F}_N^{\mathrm{o}}$: off-diagonal part of $\hat{F}_N$. $\hat{W}_N^\prime$: electron-electron repulsion term shifted by $E_{\mathrm{corr}}^{(0)}$. $\bar{F}_N$: scaled Fock operator by $\frac{1}{\langle \rm pCCD | \rm pCCD \rangle} \approx \frac{1}{ 1+\sum_{ia} |t_i^a|^2}$. The computational scaling is given in the last column. For more details see Ref.~\citenum{ap1rog-ptx}.}\label{tab:pt}
%\vspace{.2cm}
\centering\footnotesize
\begin{tabular}{lccccc}
\hline\hline
Model & $\hat{H}_0$ & $\hat{V}$ & $\langle \tilde{\Psi} \vert$ & $\hat{T}$ & scaling \\ \hline
PT2SDd & $\hat{F}_N^{\mathrm{d}}$ & $\hat{F}_N^{\mathrm{o}} + \hat{W}_N^\prime$ & $\langle 0 \vert $ & $\hat{T}_2^\prime$, $\hat{T}_1+\hat{T}_2^\prime$ & $\mathcal{O}(o^2v^2)$\\
PT2MDd & $\hat{F}_N^{\mathrm{d}}$ & $\hat{F}_N^{\mathrm{o}} + \hat{W}_N^\prime$ & $\langle \mathrm{pCCD} \vert $ & $\hat{T}_2^\prime$, $\hat{T}_1+\hat{T}_2^\prime$ & $\mathcal{O}(o^2v^2)$\\
PT2SDo & $\hat{F}_N^{\mathrm{d}} + \hat{F}_N^{\mathrm{o}}$ & $\hat{W}_N^\prime$ & $\langle 0 \vert $ & $\hat{T}_2^\prime$, $\hat{T}_1+\hat{T}_2^\prime$ & $\mathcal{O}(o^2v^3)$\\
PT2MDo & $\hat{F}_N^{\mathrm{d}} + \hat{F}_N^{\mathrm{o}}$ & $\hat{F}_N - \bar{F}_N + \hat{W}_N^\prime$ & $\langle \mathrm{pCCD} \vert $& $\hat{T}_2^\prime$, $\hat{T}_1+\hat{T}_2^\prime$ & $\mathcal{O}(o^2v^3)$\\
\multirow{2}{*}{PT2b}   & \multirow{2}{*}{$\hat{F}_N^{\mathrm{d}} + \hat{F}_N^{\mathrm{o}}$} & \multirow{2}{*}{$\hat{H}_N^\prime$} & \multirow{2}{*}{$\langle \mathrm{pCCD} \vert $}& $\hat{T}_2$,                     $\hat{T}_1+\hat{T}_2$                               &  \multirow{2}{*}{$\mathcal{O}(o^2v^3)$}\\
       &                                                   &                    &                               &              $\hat{T}_2^\prime$,                        $\hat{T}_1+\hat{T}_2^\prime$ &                       \\
\hline\hline
\end{tabular}
\end{center}
\end{table} 
%%%%%%%%%%%%%%%%%%%%%%%%%%%%%%%%%%%%%%%%%%%%%%%%%%%%%%%%%%%%%%%%%%%%%%%%%%%%%%%%%%%%%%%%%%%%%%%%%%%%%%%%%%%%%%%%%%%%%%%%%%
% End of Table 1
%%%%%%%%%%%%%%%%%%%%%%%%%%%%%%%%%%%%%%%%%%%%%%%%%%%%%%%%%%%%%%%%%%%%%%%%%%%%%%%%%%%%%%%%%%%%%%%%%%%%%%%%%%%%%%%%%%%%%%%%%%
%%%%%%%%%%%%%%%%%%%%%%%%%%%%%%%%%%%%%%%%%%%%%%%%%%%%%%%%%%%%%%%%%%%%%%%%%%%%%%%%%%%%%%%%%%%%%%%%%%%%%%%%%%%%%%%%%%%%%%%%%%
\subsubsection{A linearized coupled cluster correction with a pCCD reference function}
%%%%%%%%%%%%%%%%%%%%%%%%%%%%%%%%%%%%%%%%%%%%%%%%%%%%%%%%%%%%%%%%%%%%%%%%%%%%%%%%%%%%%%%%%%%%%%%%%%%%%%%%%%%%%%%%%%%%%%%%%%

A different approach to include dynamic correlation effects \textit{a posteriori} is to use a coupled cluster ansatz with a pCCD reference function,
%%%%%%%%%%%%%%%%%%%%%%%%%%%%%%%%%%%%%%%%%%%%%%%%%%%%%%%%%%%%%%%%%%%%%%%%%%%%%%%%%%%%%%%%%%%%%%%%%%%%%%%%%%%%%%%%%%%%%%%%%%
        \begin{equation}\label{eq:lcc}
            |\Psi \rangle = \exp({\hat{T}})  \vert {\rm pCCD} \rangle.
        \end{equation}
%%%%%%%%%%%%%%%%%%%%%%%%%%%%%%%%%%%%%%%%%%%%%%%%%%%%%%%%%%%%%%%%%%%%%%%%%%%%%%%%%%%%%%%%%%%%%%%%%%%%%%%%%%%%%%%%%%%%%%%%%%
Specifically for pCCD, the cluster operator is chosen to contain single excitations $\hat{T}_1$ and (non-pair) double excitations $\hat{T}_2^\prime$ with respect to the reference determinant of pCCD.
In the linearized CC (LCC) correction, the cluster amplitudes $t_\nu$ are determined by solving a linear set of coupled equations 
%%%%%%%%%%%%%%%%%%%%%%%%%%%%%%%%%%%%%%%%%%%%%%%%%%%%%%%%%%%%%%%%%%%%%%%%%%%%%%%%%%%%%%%%%%%%%%%%%%%%%%%%%%%%%%%%%%%%%%%%%%
        \begin{equation}
            \langle \Phi_\nu \vert (\hat{H} + [\hat{H},\hat{T}]) \vert {\rm pCCD} \rangle =  0,
        \end{equation}
%%%%%%%%%%%%%%%%%%%%%%%%%%%%%%%%%%%%%%%%%%%%%%%%%%%%%%%%%%%%%%%%%%%%%%%%%%%%%%%%%%%%%%%%%%%%%%%%%%%%%%%%%%%%%%%%%%%%%%%%%%
where the Baker--Campbell--Hausdorff expansion $e^{-\hat{T}}\hat{H}e^{\hat{T}}  = \hat{H} + [\hat{H},\hat{T}] + \frac{1}{2} [ [\hat{H},\hat{T}], \hat{T}] + \ldots$ has been truncated after the second term.
If we exploit the exponential form of the pCCD wavefunction, we can rewrite the above equation as
%%%%%%%%%%%%%%%%%%%%%%%%%%%%%%%%%%%%%%%%%%%%%%%%%%%%%%%%%%%%%%%%%%%%%%%%%%%%%%%%%%%%%%%%%%%%%%%%%%%%%%%%%%%%%%%%%%%%%%%%%%
        \begin{equation}\label{eq:pccdlcc}
            \langle \Phi_\nu \vert (\hat{H} + [\hat{H},\hat{T}] + [\hat{H},\hat{T}_{\rm p}] + [ [\hat{H},\hat{T}],\hat{T}_{\rm p}]) \vert 0 \rangle =  0.
        \end{equation}
%%%%%%%%%%%%%%%%%%%%%%%%%%%%%%%%%%%%%%%%%%%%%%%%%%%%%%%%%%%%%%%%%%%%%%%%%%%%%%%%%%%%%%%%%%%%%%%%%%%%%%%%%%%%%%%%%%%%%%%%%%
Note that, by construction, $\hat{T}$ contains excitations beyond electron pairs, which are accounted for in $\hat{T}_{\rm p}$.
Thus, the hybrid pCCD-LCCSD approach represents a simplification of CCSD and fpCCSD~\cite{frozen-pccd}, where non-pair amplitudes are treated linearly.
In contrast to conventional LCC, where the last term of Eq.~\eqref{eq:pccdlcc} is missing, the $\hat{T}_{\rm p}$ operator introduces non-linear terms, which explicitly couple the pair amplitudes with all non-pair amplitudes.
The energy can be calculated by projecting against the reference determinant of $\vert {\rm pCCD} \rangle$,
%%%%%%%%%%%%%%%%%%%%%%%%%%%%%%%%%%%%%%%%%%%%%%%%%%%%%%%%%%%%%%%%%%%%%%%%%%%%%%%%%%%%%%%%%%%%%%%%%%%%%%%%%%%%%%%%%%%%%%%%%%
        \begin{equation}
          \langle 0 \vert (\hat{H} + [\hat{H},\hat{T}]) \vert {\rm pCCD} \rangle = E,
        \end{equation}
%%%%%%%%%%%%%%%%%%%%%%%%%%%%%%%%%%%%%%%%%%%%%%%%%%%%%%%%%%%%%%%%%%%%%%%%%%%%%%%%%%%%%%%%%%%%%%%%%%%%%%%%%%%%%%%%%%%%%%%%%%
or equivalently,
%%%%%%%%%%%%%%%%%%%%%%%%%%%%%%%%%%%%%%%%%%%%%%%%%%%%%%%%%%%%%%%%%%%%%%%%%%%%%%%%%%%%%%%%%%%%%%%%%%%%%%%%%%%%%%%%%%%%%%%%%%
        \begin{equation}
          \langle 0 \vert  (\hat{H} + [\hat{H},\hat{T}] + [\hat{H},\hat{T}_{\rm p}] + [ [\hat{H},\hat{T}],\hat{T}_{\rm p}]) \vert 0 \rangle = E.
        \end{equation}
%%%%%%%%%%%%%%%%%%%%%%%%%%%%%%%%%%%%%%%%%%%%%%%%%%%%%%%%%%%%%%%%%%%%%%%%%%%%%%%%%%%%%%%%%%%%%%%%%%%%%%%%%%%%%%%%%%%%%%%%%%

%%%%%%%%%%%%%%%%%%%%%%%%%%%%%%%%%%%%%%%%%%%%%%%%%%%%%%%%%%%%%%%%%%%%%%%%%%%%%%%%%%%%%%%%%%%%%%%%%%%%%%%%%%%%%%%%%%%%%%%%%%
\subsection{Computation of interaction energies}
%%%%%%%%%%%%%%%%%%%%%%%%%%%%%%%%%%%%%%%%%%%%%%%%%%%%%%%%%%%%%%%%%%%%%%%%%%%%%%%%%%%%%%%%%%%%%%%%%%%%%%%%%%%%%%%%%%%%%%%%%%
\label{eintsection}
The main goal of the present work is to assess whether the post-pCCD methods are suitable to model systems bound by non-covalent forces. 
To this end, we will use the supermolecular approach that allows us to conveniently calculate interaction energies within any given quantum chemistry method, provided it is size-consistent.
Specifically, in the supermolecular approach the interaction energy is defined as  
\begin{align}
	E^{\rm int}_{\rm X}  = E^{\rm AB}_{\rm X} - (E^{\rm A}_{\rm X} + E^{\rm B}_{\rm X}),
	\label{eq:eint_supermolecular}
\end{align}
where the superscripts A and B denote the interacting subsystems, AB corresponds to the dimer molecule, and X indicates the employed method (e.g., $E^{\rm int}_{\rm HF}$ corresponds to the interaction energy calculated within the supermolecular Hartree--Fock approach).

It is well-known~\cite{chalasinski2000,counter-poise-correction} that such a subtraction must be performed with caution.
%as the subtracted quantities are very large and the final interaction energy can be smaller than the corresponding total energies by a few orders of magnitude.
To this end, total energies of the monomers and of the dimer had been obtained consistently, that is, within exactly the same methods, basis sets, geometries, and numerical thresholds of accuracy.

%One should also keep in mind that the final results might be prone to the so-called basis set superposition error (BSSE)~\cite{}. 
%It might happen that the dimer energy and the sum of the monomer energies are different, resulting in a better description of the monomer energies around the equilibrium geometry compared with stretched intermonomer distances.  
%To minimize the BSSE effect, all terms on the right hand side of Eq.~\eqref{eq:eint_supermolecular} should be calculated in a dimer-centered basis. 

To probe the composition of the pCCD and post-pCCD interaction energies and to better understand the role of each component in forming different types of non-covalent compounds, we will utilize the interaction energy decomposition provided within the SAPT framework.
In the many-electron formulation of SAPT~\cite{jeziorski1994,rybak1991,moszynski1994,szalewicz2012} the interaction energy is obtained directly, with no subtraction whatsoever.
Instead, it is obtained as the sum of the lowest corrections in the perturbation series, in which the system Hamiltonian is partitioned into the intermonomer interaction $V_{\rm AB}$ and the Hamiltonians corresponding to the monomers $\hat{H}_{\rm A}$ and $\hat{H}_{\rm B}$, 
\begin{equation}
	\hat{H}=\hat{H}_{\rm A}+\hat{H}_{\rm B} +  \hat{V}_{\rm AB}.
\end{equation}
The $ \hat{V}_{\rm AB}$ operator includes all Coulomb repulsions between the electrons of monomer A and B, the repulsion between the nuclei of A and B, and all attractive interactions between the electrons of monomer A and the nuclei of monomer B, and {\em  vice versa}.
Such a partitioning of the Hamiltonian defines the perturbation series in terms of the $\lambda$ parameter.
The zeroth-order wavefunction in SAPT is the product of the wavefunctions of the noninteracting monomers. 
Furthermore, in the SAPT decomposition of interaction energies, it is well known that the energy expressions of arbitrary order have to be modified to ensure the proper permutation symmetry of all electrons within the dimer.
This procedure (see, for example, Ref.~\citenum{jeziorski1978symmetry}) results in the appearance of the so-called exchange corrections in the perturbation series, which is required for a proper electron exchange between the monomers.
Nowadays, the many-electron SAPT approach is a well-establish theory and routinely applied. 
For practical reasons, however, only the lowest two orders in $\hat{V}_{\rm AB}$ are used to calculate the interaction energies,
\begin{equation}
	E_{\rm int}^{\rm SAPT } = E^{(1)}_{\rm elst} + E^{(1)}_{\rm exch} + E^{(2)}_{\rm ind}+ E^{(2)}_{\rm exch-ind} + 
	 E^{(2)}_{\rm disp}+ E^{(2)}_{\rm exch-disp}. 
	\label{saptcorrect}
\end{equation}
The first two corrections on the right hand side of the above equation represent the electrostatic interaction energy of molecular charge distributions and the correction for the exchange energy, which is the dominant repulsive effect.
 $E^{(2)}_{\rm ind}$ is the induction energy and originates from the mutual polarization of the monomers in the field of their corresponding partners, while $E^{(2)}_{\rm exch-ind}$ is the exchange--induction energy.
The $E^{(2)}_{\rm disp}$ and $E^{(2)}_{\rm exch-disp}$ terms are the dispersion and exchange--dispersion energies~\cite{chalasinski1976,chalasinski1977,korona2008,korona2009} and are related to the correlated motion of electrons between the monomers. 

In mean-field methods such as Hartree--Fock theory or the CASSCF approach with a minimal active space, dynamic electron correlation is missing, which often results in catastrophic errors when predicting interaction energies.
Most strikingly in such calculations is that the dispersion energy is missing.~\cite{jeziorski1976variation,moszynski1996sapt,hepburn1975,jeziorska1987}
Thus in dispersion dominated systems, like systems with non-polar molecules or atoms, this deficiency might even lead to the wrong sign of the interaction energy.
On the other hand, mean-field methods correctly reproduce the induction and electrostatic energies as well as higher-order induction effects.~\cite{jeziorski1976variation,moszynski1994ind}

It is, however, not entirely clear how our geminal-based methods perform for interaction energies of non-covalent systems. 
In order to reproduce the dispersion energy in van der Waals complexes, it is essential to include \textit{intersystem} excitations with at least one single excitation on \textit{each} subsystem.
In the (restricted) pCCD ansatz (cf. Eq.~\eqref{eq:ap1rog}), the $\hat{T}_\textrm{p}$ operator is by construction a double excitation operator of opposite spin pairs, but bearing the same spatial parts.  Hence, it is not possible for the dimer $|\textrm{pCCD}\rangle$ wavefunction to reproduce the dispersion energy.
Recently Garza \textit{et al.}\mbox{~\citenum{garza_rpa}} made a similar observation for the helium and neon dimers (for these systems pCCD yields repulsive interaction energies), which confirms the inability of pCCD to reproduce dispersion interactions.

On the other hand, orbital-optimized pCCD is exact for two-electron systems. Hence, the intramonomer correlation energy should be reproduced to some extent.  A comparison between the supermolecular pCCD and dispersion-free SAPT (labeled from now on as SAPT2df) interaction energies and the supermolecular Hartree--Fock energy will shed some light on this puzzle.
Throughout this paper, we will consider dispersion-free SAPT as a sum of the following terms
\begin{widetext}
	\begin{equation}\label{eq:sapt_disp_free}
		E_{\rm{SAPT2df}} = E_{\rm{SAPT2}} - E^{(20)}_{\rm disp} - E^{(20)}_{\rm exch-disp},
	\end{equation}
\end{widetext}
where SAPT2 is defined as
\begin{widetext}
	\begin{equation}\label{eq:sapt2}
		E_{\rm{SAPT2}} = E_{\rm{SAPT0}} + E^{(12)}_{\rm elst,resp} +  E^{(11)}_{\rm exch} + E^{(12)}_{\rm exch} + \,^{t}E^{(22)}_{\rm ind} + \,^{t}E^{(20)}_{\rm exch-ind}, 
	\end{equation}
\end{widetext}
    with
\begin{widetext}
	\begin{equation}\label{eq:sapt0}
		E_{\rm{SAPT0}} = E^{(10)}_{\rm elst} +  E^{(10)}_{\rm exch} +  E^{(20)}_{\rm ind,resp} + E^{(20)}_{\rm exch-ind,resp} + E^{(20)}_{\rm disp} + E^{(20)}_{\rm exch-disp} +  \delta^{(2)}_{\rm HF}.
	\end{equation}
\end{widetext}
Thus, dispersion-free SAPT accounts for intramonomer correlation through electrostatics and exchange in first order as well as induction in second order.
It also includes infinite-order induction effects by adding the so-called $\delta^{(2)}_{\rm HF}$ term, which is the difference between $ E^{(10)}_{\rm elst} + E^{(10)}_{\rm exch} + E^{(20)}_{\rm ind,resp} + E^{(20)}_{\rm exch-ind,resp}$  and the supermolecular  Hartree--Fock interaction energy. 
The superscript \textit{t} in Eq.~\eqref{eq:sapt2} indicates the ``true'' correlation contribution. 
We refer the reader to Ref.~\citenum{jeziorski1994} for more details concerning eqs.~\eqref{eq:sapt2} and~\eqref{eq:sapt0}.  

The supermolecular pCCD-LCCSD interaction energy includes double excitations that occupy different spatial orbitals between the systems.
It is, thus, quite obvious that the dispersion energy should be properly recovered within this method.
However, it is well known that dispersion can be very sensitive to the quality of the monomer wavefunction~\cite{rybak1991}.
In particular, the contribution of triply excited configurations is often crucial and  can contribute to even 30\% of the total interaction energy~\cite{zuchowski2014,rajchel2007,patkowski2007metals}.
In this paper, we will compare this LCCSD correction to the closed-shell supermolecular CEPA(0) (or equivalently linearized CCSD) and CCSD interaction energies for the A24 set to assess the performance of pCCD-LCCSD. 

Finally, we expect that the pCCD-based perturbation theories can reproduce dispersion energies similar to M\o{}ller--Plesset perturbation theory of second order (MP2). %just like supermolecular interaction energy in MP2 model do.
As it is well-known, the MP2 interaction energy in the supermolecular approach includes the $E^{(20)}_{\rm disp}$ interaction energy  (with exchange counterpart), where the second number in the superscript refers to the correction in terms of the intramonomer correlation~\cite{chalasinski1988,korona2008,korona2009}.
Such an approximation of the dispersion energy is equivalent to a Hartree--Fock description of the monomers, which can be very poor in some cases, like the interaction of $\pi$-stacked aromatic molecules~\cite{hesselmann2011,pitonak2010}. 

Since the corrections to pCCD are designed to describe both static and dynamic electron correlation, we will carefully examine the interaction energies in systems where the multi-reference character of the wavefunction directly affects the interaction energy.
We expect pCCD to correctly reproduce static correlation associated with the multi-configurational description of the monomers, while the LCCSD and PT models---as mentioned above---properly recover the dispersion energy. 

% The first such system is H$_2\cdots$H$_2$ in T-shape configuration in  which the hydrogen which points the center of mass of other molecule is being stretched, so that the interaction energy limit for the dissociation of H$_2$ correspond to  H$_2\cdots $H system.
%The LiH$\cdots$H$_2$ dimer in which LiH is stretched will show somewhat similar effect but for the case of heteronuclear molecule, so it is possible that the system dissociates into ion-pair  Li$^+$H$^-$  limit in case of the Hartree--Fock theory of into neutral Li$\cdots$H system.

%%%%%%%%%%%%%%%%%%%%%%%%%%%%%%%%%%%%%%%%%%%%%%%%%%%%%%%%%%%%%%%%%%%%%%%%%%%%%%%%%%%%%%%%%%%%%%%%%%%%%%%%%%%%%%%%%%%%%%%%%%
%%%%%%%%%%%%%%%%%%%%%%%%%%%%%%%%%%%%%%%%%%%%%%%%%%%%%%%%%%%%%%%%%%%%%%%%%%%%%%%%%%%%%%%%%%%%%%%%%%%%%%%%%%%%%%%%%%%%%%%%%%
\section{Computational Details}\label{sec:comp}
%%%%%%%%%%%%%%%%%%%%%%%%%%%%%%%%%%%%%%%%%%%%%%%%%%%%%%%%%%%%%%%%%%%%%%%%%%%%%%%%%%%%%%%%%%%%%%%%%%%%%%%%%%%%%%%%%%%%%%%%%%
%%%%%%%%%%%%%%%%%%%%%%%%%%%%%%%%%%%%%%%%%%%%%%%%%%%%%%%%%%%%%%%%%%%%%%%%%%%%%%%%%%%%%%%%%%%%%%%%%%%%%%%%%%%%%%%%%%%%%%%%%%
\subsection{Basis sets and extrapolation to the basis set limit}
%%%%%%%%%%%%%%%%%%%%%%%%%%%%%%%%%%%%%%%%%%%%%%%%%%%%%%%%%%%%%%%%%%%%%%%%%%%%%%%%%%%%%%%%%%%%%%%%%%%%%%%%%%%%%%%%%%%%%%%%%%
For all atoms, we considered the augmented correlation consistent series of basis sets developed by Dunning and coworkers.  
%To minimize the basis set superposition error in supermolecular calculations, the mid bond functions of Sl{\'{a}}dek
%\textit{et al.}~\cite{Sladek2011} (2$s$3$p$2$d$2$f$2$g$) \filip{(Didn't really used mid bonds so far)}
%were added and the resulting energies were counter point corrected~\cite{counter-poise-correction}.
Specifically for the A24 test set (see Figure~\ref{fig:a24}), we used the aug-cc-pVDZ, aug-cc-pVTZ, and aug-cc-pVQZ basis sets~\cite{aug-cc-pvtz}. 
The corresponding auxiliary basis sets~\cite{CC-PVXZ-JKFIT,CC-PVXZ-RIFIT} for density fitting were employed for calculating all SAPT energies using the PSI4 software package~\cite{psi4-2012,psi4-2017}. 
In all pCCD-based methods, pivoted incomplete Cholesky decomposition (CD) was used to approximate the electron repulsion integral tensors.
The threshold cutoff ($\delta_{\rm CD}$) of the CD procedure was set to values ranging from 10$^{-8}$ to 10$^{-10}$ for all systems, which is more than sufficient to retrieve $\mu$E$_\text{h}$ accuracy.
Furthermore, we performed an extrapolation to the basis set limit using a two-step procedure~\cite{helgaker1997,halkier1998a}.
First, the basis set limit of the Hartree--Fock energy was obtained by fitting an exponential function of the form~\cite{helgaker1997}
\begin{equation}\label{eq:hf_extrapolation}
	E^{\rm SCF}\left(X\right) = E^{\rm SCF}_{\infty} + a\exp\left(-bX\right), 
\end{equation}
to the Hartree--Fock energies (E$^{\rm SCF}$) obtained with the aug-cc-pVDZ, aug-cc-pVTZ, and aug-cc-pVQZ basis sets.
In the above equation, $X$ indicates the cardinal number of the basis set (2 for D, 3 for T, etc.). 
Second, the correlation energies were extrapolated by a two-point fit to the function~\cite{helgaker1997, halkier1998a}
\begin{equation}\label{eq:corr_extrapolation}
	E^{\rm corr}\left(X\right) = E^{\rm corr}_{\infty} + aX^{-3}, 
\end{equation}
where $E^{\rm corr}\left(X\right)$ is defined as $E^{\rm corr}\left(X\right) = E^{\rm tot}\left(X\right) - E^{\rm SCF}\left(X\right)$.
For the correlation energy, we included only results obtained with the aug-cc-pVTZ and aug-cc-pVQZ basis sets in the fitting procedure.
In all pCCD-based methods, the Hartree--Fock energy was extrapolated via Eq.~\eqref{eq:hf_extrapolation}, then the pair correlation energy was treated using the extrapolation scheme given in Eq.~\eqref{eq:corr_extrapolation}.
All \textit{a posteriori} correlation corrections on top of pCCD were extrapolated in the same fashion. 

All other calculations were performed using the aug-cc-pVDZ and aug-cc-pVTZ~\cite{aug-cc-pvtz} basis sets if not stated otherwise. 
For all investigated molecules, we applied a counterpoise correction to minimize the basis set superposition error~\cite{counter-poise-correction} (\textit{vide infra}).  

In the case of H$_3^+\cdots$H$_2$ two types of basis sets were employed: aug-cc-pVDZ for 2-electron full configuration interaction(FCI)-based SAPT calculations and a modified aug-cc-pVTZ, where all diffuse d-type functions were neglected (see section S2 in the Supplementary Information for more details).

%%%%%%%%%%%%%%%%%%%%%%%%%%%%%%%%%%%%%%%%%%%%%%%%%%%%%%%%%%%%%%%%%%%%%%%%%%%%%%%%%%%%%%%%%%%%%%%%%%%%%%%%%%%%%%%%%%%%%%%%%%
\subsection{pCCD and pCCD-based corrections}
%%%%%%%%%%%%%%%%%%%%%%%%%%%%%%%%%%%%%%%%%%%%%%%%%%%%%%%%%%%%%%%%%%%%%%%%%%%%%%%%%%%%%%%%%%%%%%%%%%%%%%%%%%%%%%%%%%%%%%%%%%
All pCCD-based calculations were performed using our own software package \textsc{Piernik}~\cite{piernik100}.
In the variational orbital optimization pCCD calculations, the total energy convergence threshold was set to $10^{-11}$ $E_\text{h}$ (or tighter) with a geminal coefficients threshold of 10$^{-12}$ for the absolute tolerance in the maximum norm for the residual (or tighter), while the absolute tolerance in the maximum norm of the orbital gradient was set to 10$^{-6}$ (or tighter).
For the LCCSD and all PT2 corrections with a pCCD reference function (in the optimized orbital basis), the convergence threshold for the energy was set to 10$^{-10}$ $E_\text{h}$  (or tighter), while the absolute tolerance in the maximum norm of the amplitudes residue was set to 10$^{-9}$ (or tighter).

% Now we will investigate how the pCCD and post-pCCD methods reproduce the interaction energies and to what extent they can be used for supermolecular calculations of potential energy surfaces.
One of the most important issues in supermolecular calculations is to ensure that both the dimer and monomers converge to consistent states during the optimization procedure. To this end, we monitored the orbitals and their occupancies in both monomer and dimer calculations, while for the dissociating monomers we employed the orbitals from an adjacent point along the dissociation pathway as initial guess orbitals (specifically, moving towards decreasing monomer separations). 

%%%%%%%%%%%%%%%%%%%%%%%%%%%%%%%%%%%%%%%%%%%%%%%%%%%%%%%%%%%%%%%%%%%%%%%%%%%%%%%%%%%%%%%%%%%%%%%%%%%%%%%%%%%%%%%%%%%%%%%%%%
\subsection{Reference methods}
%%%%%%%%%%%%%%%%%%%%%%%%%%%%%%%%%%%%%%%%%%%%%%%%%%%%%%%%%%%%%%%%%%%%%%%%%%%%%%%%%%%%%%%%%%%%%%%%%%%%%%%%%%%%%%%%%%%%%%%%%%

All CEPA(0), CCSD, and SAPT calculation for the A24 test set were performed with the \textsc{Psi4} software package~\cite{psi4-2012,psi4-2017}.
In all calculations, we applied canonical restricted Hartree--Fock (RHF) orbitals.

The SAPT calculations included intramonomer electronic correlation as described in refs.~\citenum{moszynski1994,rybak1991,jeziorski1994}. The density-fitting employed in all SAPT calculations used JK-fit and RI-fit auxiliary basis sets corresponding to the aug-cc-pVXZ (X = 2, 3, 4) primary basis set. 
The SAPT2df interaction energy was obtained using Eq. \eqref{eq:sapt_disp_free}.

For H$_3^{+}\cdots$H$_2$ 2-electron sub-systems, we performed SAPT calculations based on a FCI wavefunction.
To this end, we used the SAPT2012 package~\cite{sapt2012} (with the \textsc{Molpro} program) to generate the molecular integrals, which was later interfaced with the computer code utilized in Refs.~\citenum{korona_4el,williams1996,korona1997}.
We should emphasize that this code allows us to consider only small basis sets, not larger than approximately 80 basis functions.

For LiH$\cdots$H$_2$, we also performed CASSCF and multi-reference configuration interaction singles and doubles (MRCI-SD) calculations.
In order to correctly reproduce the dissociation limit of the LiH molecule, we included only the lowest antibonding orbital into the active space so that the active space contains the $1s$ orbital of Li, the $\sigma$ and $\sigma^*$ orbitals of LiH, and the $\sigma(1s)$ orbital on H$_2$.
%We will denote the calculations with such active space as to CASSCF and MRCISD, respectively.
We also performed FCI calculations for LiH with a frozen $1s$ shell for the Li atom.
The energy convergence thresholds for FCI, CASSCF, and MRCI-SD were all set to 10$^{-7}$ E$_\text{h}$.  
Finally, we should note that we encountered convergence difficulties in CEPA(0) calculations for molecular systems with stretched bonds or near conical intersections. Thus, the corresponding CEPA(0) results are not shown.

%%%%%%%%%%%%%%%%%%%%%%%%%%%%%%%%%%%%%%%%%%%%%%%%%%%%%%%%%%%%%%%%%%%%%%%%%%%%%%%%%%%%%%%%%%%%%%%%%%%%%%%%%%%%%%%%%%%%%%%%%%
%%%%%%%%%%%%%%%%%%%%%%%%%%%%%%%%%%%%%%%%%%%%%%%%%%%%%%%%%%%%%%%%%%%%%%%%%%%%%%%%%%%%%%%%%%%%%%%%%%%%%%%%%%%%%%%%%%%%%%%%%%
\section{Results and Discussion}\label{sec:results}
%%%%%%%%%%%%%%%%%%%%%%%%%%%%%%%%%%%%%%%%%%%%%%%%%%%%%%%%%%%%%
%%%%%%%%%%%%%%%%%%%%%%%%%%%%%%%%%%%%%%%%%%%%%%%%%%%%%%%%%%%%%%%%%%%%%%%%%%%%%%%%%%%%%%%%%%%%%%%%%%%%%%%%%%%%%%%%%%%%%%%%%%
%%%%%%%%%%%%%%%%%%%%%%%%%%%%%%%%%%%%%%%%%%%%%%%%%%%%%%%%%%%%%%%%%%%%%%%%%%%%%%%%%%%%%%%%%%%%%%%%%%%%%%%%%%%%%%%%%%%%%%%%%%
%%%%%%%%%%%%%%%%%%%%%%%%%%%%%%%%%%%%%%%%%%%%%%%%%%%%%%%%%%%%%%%%%%%%%%%%%%%%%%%%%%%%%%%%%%%%%%%%%%%%%%%%%%%%%%%%%%%%%%%%%%
\begin{figure*}[tb]
 \centering
 \includegraphics[width=0.5\textwidth]{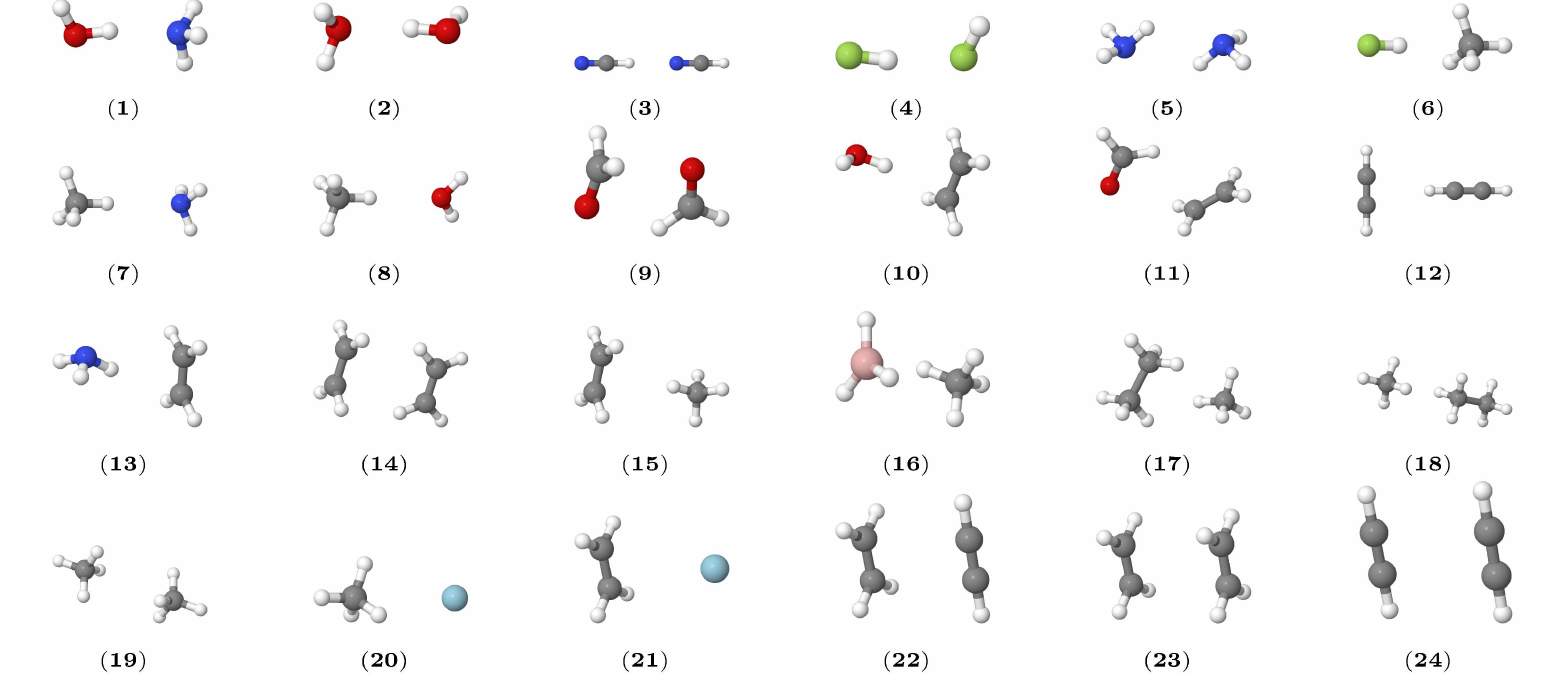}
 \caption{\footnotesize Molecular structures of all molecules contained in the A24 set.
(1) \ce{H2O\cdots NH3},
(2) \ce{H2O\cdots H2O},						
(3) \ce{HCN\cdots HCN},					
(4) \ce{HF\cdots HF},							
(5) \ce{NH3\cdots NH3},						
(6) \ce{HF\cdots CH4},				
(7) \ce{NH3\cdots CH4},		
(8) \ce{H2O\cdots CH4},
(9) \ce{HCHO\cdots HCHO},
(10) \ce{H2O\cdots C2H4},
(11) \ce{HCHO\cdots C2H4},
(12) \ce{C2H2\cdots C2H2} (T-shaped),
(13) \ce{NH3\cdots C2H4},
(14) \ce{C2H4\cdots C2H4} (T-shaped),
(15) \ce{CH4\cdots C2H4},
(16) \ce{BH3\cdots CH4},
(17) \ce{CH4\cdots C2H6} (side-on),
(18) \ce{CH4\cdots C2H6} (end-on),
(19) \ce{CH4\cdots CH4},
(20) \ce{Ar\cdots CH4},
(21) \ce{Ar\cdots C2H4},
(22) \ce{C2H4\cdots C2H2},
(23) \ce{C2H4\cdots C2H4} (parallel),
(24) \ce{C2H2\cdots C2H2} (parallel).
    }\label{fig:a24}
\end{figure*}
%%%%%%%%%%%%%%%%%%%%%%%%%%%%%%%%%%%%%%%%%%%%%%%%%%%%%%%%%%%%%%%%%%%%%%%%%%%%%%%%%%%%%%%%%%%%%%%%%%%%%%%%%%%%%%%%%%%%%%%%%%
%%%%%%%%%%%%%%%%%%%%%%%%%%%%%%%%%%%%%%%%%%%%%%%%%%%%%%%%%%%%%%
\subsection{ A24 data set}
%%%%%%%%%%%%%%%%%%%%%%%%%%%%%%%%%%%%%%%%%%%%%%%%%%%%%%%%%%%%%%%%%%%%%%%%%%%%%%%%%%%%%%%%%%%%%%%%%%%%%%%%%%%%%%%%%%%%%%%%%%

\begin{figure}[tb]
	\centering
	\includegraphics[width=.5\textwidth]{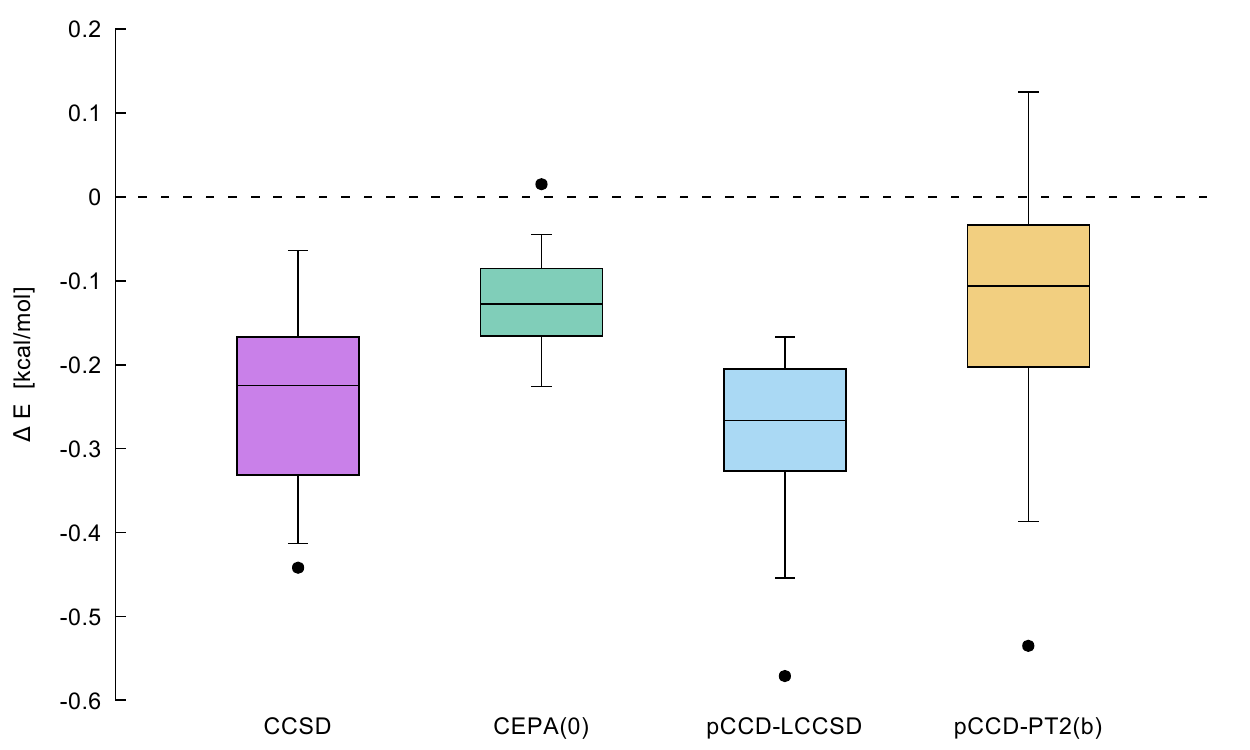}
	\caption{Box plot of errors [kcal/mol] obtained with some selected methods (for numerical values see Table~\ref{tab:a24}). All errors are given with respect to the CCSD(T)/CBS reference. The box represents the likely range of variation (interquartile range), while the outer fences cover 95\% of the results.}
	\label{fig:a24_errors}
\end{figure}

Our first test system to assess the accuracy of the above mentioned pCCD-based methods in describing non-covalent interactions contains the A24 test set~\cite{a24} as shown in Figure~\ref{fig:a24}.
Table~\ref{tab:a24} summarizes all interaction energies obtained by various coupled cluster methods (CCSD, CEPA(0), pCCD including a dynamic energy correction, and CCSD(T)) extrapolated to the basis set limit (CBS).
All error measures mentioned in the Table are given with respect to CCSD(T) reference data.~\cite{a24}
We should emphasize that only the best-performing PT2 correction is shown in Table~\ref{tab:a24}, i.e., PT2b(d).
Note, however, that the whole PT2b family yields similar results. 
All remaining PT2 corrections (see Table~\ref{tab:pt}) result in larger errors (greater or equal to 1 kcal/mol) and are briefly summarized in Table S1 of the Supporting Information.
Note that the conventional CC calculations (CCSD, CEPA(0), and CCSD(T)) are performed within canonical RHF orbitals, while the orbital basis has been optimized for pCCD-based methods.
In general, all coupled cluster methods restricted to at most double excitations underestimate the (extrapolated) interaction energies (see Figure~\ref{fig:a24_errors} as well as Table~\ref{tab:a24} for more details).
The performance of CEPA(0) is the best and can be understood from the fact that the disconnected quadruple diagrams, which contribute to the dispersion energy are repulsive~\cite{moszynski1994} and they are neglected in linearized CCSD theory.
Hence, the overall interaction energy in linearized CCSD (which is equivalent to CEPA(0)) is slightly more attractive. 
Although our pCCD-LCCSD interaction energies are slightly more repulsive than the CCSD interaction energies, they feature tighter outer fences representing 95 $\%$ (see Figure~\ref{fig:a24_errors}) of the A24 test set with one outliner (system no.~9) around $-$0.57 kcal/mol.   
We should stress that pCCD-PT2b(d) outperforms both pCCD-LCCSD and CCSD, reducing the MAE and RMSE by roughly a factor of 2.
However, the (averaged) good performance of pCCD-PT2b(d) comes at the price of a wider variation compared to all tested quantum chemistry methods (see also Figure~\ref{fig:a24_errors}).
Specifically, while for some systems the pCCD-PT2b(d) interaction energy is almost identical to the CCSD(T) reference energy (\textit{e.g.}, system 24 in Table \ref{tab:a24}), a PT2b correction underestimates the interaction energy by over 50 \% for other test systems, like BH$_3\cdots$CH$_4$ (see also Table~\ref{tab:a24}).
Finally, we should stress that the PT2b variants perform best compared to other perturbative approaches on top of pCCD.  Similar observations have been made for reaction energies as reported in Ref.~\citenum{ap1rog-ptx}.

\begin{table*}[tb]
\begin{center}
	\caption{A24 interaction energies in (kcal/mol) extrapolated to CBS limit. Differences with respect to CCSD(T) are given in parenthesis. The molecular structures of all systems are shown in Figure~\ref{fig:a24}. ME: mean error, MAE: mean absolute error, RMSE: root mean squared error, max AE: maximal absolute error. 
    }\label{tab:a24}
	\scriptsize
	\begin{tabular}{c r@.l r@.l r@.l r@.l r@.l}
	\hline \hline
		System & \mc{2}{c}{CCSD(T)/CBS} & \mc{2}{c}{CCSD/CBS} & \mc{2}{c}{CEPA(0)/CBS} & \mc{2}{c}{pCCD-LCCSD/CBS} &\mc{2}{c}{pCCD-PT2b(d)/CBS}\\\hline
		1  &\phantom{000}$-$6&493 &\phantom{00}$-$6&210($-$0.324) &\phantom{00}$-$6&409($-$0.125) &\phantom{00}$-$6&367($-$0.167) &\phantom{00}$-$6&649($+$0.115) \\
		2  &             $-$5&006 &            $-$4&859($-$0.179) &            $-$4&993($-$0.045) &            $-$4&629($-$0.409) &            $-$4&959($-$0.079) \\
		3  &             $-$4&745 &            $-$4&648($-$0.115) &            $-$4&585($-$0.178) &            $-$4&431($-$0.332) &            $-$4&684($-$0.079) \\
		4  &             $-$4&581 &            $-$4&542($-$0.064) &            $-$4&621($+$0.015) &            $-$4&420($-$0.186) &            $-$4&509($-$0.097) \\
		5  &             $-$3&137 &            $-$2&908($-$0.246) &            $-$3&064($-$0.090) &            $-$2&844($-$0.310) &            $-$3&031($-$0.123) \\
		6  &             $-$1&654 &            $-$1&366($-$0.308) &            $-$1&516($-$0.158) &            $-$1&309($-$0.365) &            $-$1&287($-$0.387) \\
		7  &             $-$0&765 &            $-$0&615($-$0.158) &            $-$0&700($-$0.073) &            $-$0&495($-$0.278) &            $-$0&632($-$0.141) \\
		8  &             $-$0&663 &            $-$0&469($-$0.200) &            $-$0&534($-$0.135) &            $-$0&391($-$0.278) &            $-$0&472($-$0.197) \\
		9  &             $-$4&554 &            $-$4&145($-$0.413) &            $-$4&351($-$0.207) &            $-$3&987($-$0.571) &            $-$4&286($-$0.272) \\
		10 &             $-$2&557 &            $-$2&233($-$0.338) &            $-$2&345($-$0.226) &            $-$2&193($-$0.378) &            $-$2&361($-$0.210) \\
		11 &             $-$1&621 &            $-$1&346($-$0.281) &            $-$1&438($-$0.189) &            $-$1&173($-$0.454) &            $-$1&443($-$0.184) \\
		12 &             $-$1&524 &            $-$1&341($-$0.188) &            $-$1&363($-$0.166) &            $-$1&298($-$0.231) &            $-$1&492($-$0.037) \\
		13 &             $-$1&374 &            $-$1&154($-$0.227) &            $-$1&244($-$0.137) &            $-$1&071($-$0.310) &            $-$1&342($-$0.039) \\
		14 &             $-$1&090 &            $-$0&777($-$0.316) &            $-$0&928($-$0.165) &            $-$0&877($-$0.216) &            $-$0&977($-$0.116) \\
		15 &             $-$0&502 &            $-$0&362($-$0.143) &            $-$0&424($-$0.081) &            $-$0&301($-$0.204) &            $-$0&475($-$0.030) \\
		16 &             $-$1&485 &            $-$1&126($-$0.363) &            $-$1&358($-$0.131) &            $-$1&167($-$0.322) &            $-$0&954($-$0.535) \\
		17 &             $-$0&827 &            $-$0&604($-$0.223) &            $-$0&730($-$0.097) &            $-$0&584($-$0.243) &            $-$0&598($-$0.229) \\
		18 &             $-$0&607 &            $-$0&440($-$0.167) &            $-$0&541($-$0.066) &            $-$0&400($-$0.207) &            $-$0&399($-$0.208) \\
		19 &             $-$0&533 &            $-$0&393($-$0.139) &            $-$0&475($-$0.057) &            $-$0&353($-$0.179) &            $-$0&367($-$0.165) \\
		20 &             $-$0&405 &            $-$0&230($-$0.177) &            $-$0&284($-$0.123) &            $-$0&215($-$0.192) &            $-$0&339($-$0.068) \\
		21 &             $-$0&364 &            $-$0&198($-$0.168) &            $-$0&247($-$0.119) &            $-$0&171($-$0.195) &            $-$0&428($+$0.062) \\
		22 &                0&821 &               1&208($-$0.398) &               0&961($-$0.151) &               1&065($-$0.255) &               0&736($+$0.074) \\
		23 &                0&934 &               1&364($-$0.442) &               1&029($-$0.107) &               1&217($-$0.295) &               0&797($+$0.125) \\
		24 &                1&115 &               1&467($-$0.363) &               1&285($-$0.181) &               1&342($-$0.238) &               1&108($-$0.004) \\
	\hline\hline                                         
				ME & \mc{2}{c}{--} & $-$0&248 & $-$0&125 & $-$0&284 & $-$0&118\\             
       MAE & \mc{2}{c}{--} &    0&248 &    0&126 &    0&284 &    0&149\\
      RMSE & \mc{2}{c}{--} &    0&268 &    0&136 &    0&300 &    0&190\\
    max AE & \mc{2}{c}{--} &    0&442 &    0&226 &    0&571 &    0&535\\
	\hline\hline
\end{tabular}
\end{center}
\end{table*}

\begin{figure}[tb]
	\centering
	\includegraphics[width=0.5\textwidth]{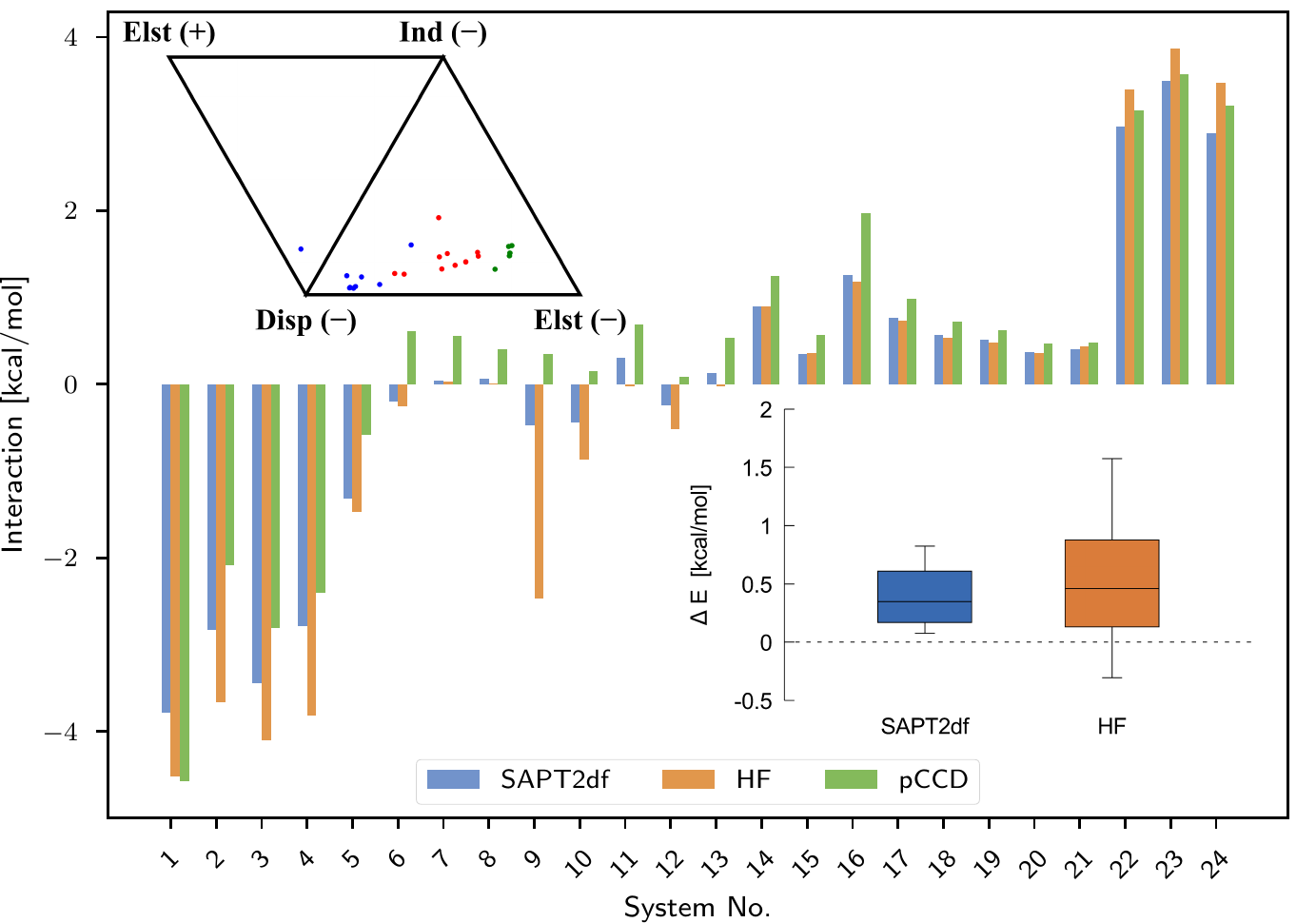}
	\caption{Bar plot of interaction energies obtained with SAPT2df, HF, and pCCD. Top-left inset: ternary diagram~\cite{ternary_singh,ternary_burns} based on SAPT0 corrections obtained with the aug-cc-pVQZ basis set. The subsets are divided according to Hobza {\em et al.}~\cite{a24} and color-coded. Green: hydrogen bonds, red: mixed interactions, blue: dispersion dominated. Bottom-right inset: box plot illustrating the differences in interaction energy between pCCD and SAPT2df or HF, respectively. The box represents the likely range of variation (interquartile range), while the outer fences cover 95\% of the results.}
	\label{fig:a24_displess_barplot}
\end{figure}

Having tested various CC and PT2 methods that include dynamic correlation effects on top of pCCD \textit{a posteriori}, we will now focus on the pCCD supermolecular interaction itself.
As we have pointed out in Sec.~\ref{eintsection}, we expect that the pCCD method does not include any dispersion interaction between two subsystems as its wavefunction ansatz allows only for pair excitations and the dispersion interaction of lowest order arises from double excitations, where each of the monomers contributes to at least one single excitation (so-called broken-pair excitations).
Although dispersion contributions in the supermolecular pCCD energy are not expected, one can argue that correlated electrostatic and induction contributions are present since pCCD includes intramonomer correlation. In particular, pCCD with optimized orbitals is exact for two-electron systems~\cite{ap1rog-jctc}.
For these reasons, we decided to compare the pCCD supermolecular interaction energy with the one obtained from supermolecular Hartree--Fock calculations, which include the uncorrelated electrostatics and induction contributions, but no dispersion, as well as from dispersion-free SAPT2 calculations (see Eq.~\ref{eq:sapt_disp_free}), which include additional contributions of correlated electrostatics and induction.
This analysis allows us to scrutinize the contributions present in supermolecular pCCD.
The corresponding data is presented as inset in Figure~\ref{fig:a24_displess_barplot}.
Statistically, the interaction predicted by pCCD (see inset of Figure~\ref{fig:a24_displess_barplot}) tends to be more repulsive than obtained within SAPT2df and HF (median of 0.55 and 0.73 kcal/mol respectively).
For systems 14-24,  which  are generally dominated by dispersion, pCCD agrees with both theories (with differences of approximately 0.2 kcal/mol), except for BH$_3\cdots$NH$_3$, where both SAPT2df and HF yield smaller interaction energies by about 0.5 kcal/mol.
Interestingly, pCCD predicts a slightly positive interaction energy for systems 6-13, where both SAPT2df and supermolecular HF yield either marginally positive or negative interaction energies.
The absolute differences between $E^{\rm int}_{\rm pCCD}$, $E^{\rm int}_{\rm HF}$, and $E^{\rm int}_{\rm SAPT2df}$ for the hydrogen-bonded systems~\cite{a24} 1-5 are larger than for the remaining systems in the A24 test set.
This might suggest that the dispersion-free interaction energy is more sensitive for systems that contain more polar molecules.
What is still missing in SAPT2df and HF is the coupling of the intramonomer correlation with the higher-order induction energy.
Clearly, if intramonomer correlation strongly affects the dipole moments or polarizabilities, the remainder of the induction energy terms higher than second order is also affected by intramonomer correlation effects.
This may lead to larger discrepancies in interaction energies between pCCD and SAPT2df and pCCD and HF, respectively.

%\begin{figure}[tb]
%	\centering
%	\includegraphics[width=1.0\textwidth]{./figs/a24/a24_displess_1.pdf}
%	\caption{Box plot illustrating the differences in interaction energy between pCCD and SAPT2df and HF, respectively. The box represents the likely range of variation (interquartile range), while the outer fences cover 95\% of the results. \filip{Will be visually overhauled}}
%	\label{fig:a24_displess_boxplot}
%\end{figure}

%%%%%%%%%%%%%%%%%%%%%%%%%%%%%%%%%%%%%%%%%%%%%%%%%%%%%%%%%%%%%%%%%%%%%%%%%%%%%%%%%%%%%%%%%%%%%%%%%%%%%%%%%%%%%%%%%%%%%%%%%%
\subsection{Systems with mixed static and dynamic correlation}
%%%%%%%%%%%%%%%%%%%%%%%%%%%%%%%%%%%%%%%%%%%%%%%%%%%%%%%%%%%%%%%%%%%%%%%%%%%%%%%%%%%%%%%%%%%%%%%%%%%%%%%%%%%%%%%%%%%%%%%%%%
In this subsection, we will investigate the interaction energies in small model systems, where one of the monomers contains stretched bonds and thus the whole system features both static and dynamic electron correlation. 
Specifically, our model test systems include (a) T-shaped H$_2\cdots$ H$_2$ with one of the monomers being stretched, (b) H$_2\cdots$ LiH with a stretched Li--H bond, and (c) the avoided crossing in H$_3^+\cdots$ H$_2$. 
Based on our previous observations for the A24 set, we will focus on the most promising post-pCCD corrections investigated in this work, that is, PT2b(d) and LCCSD.  
%%%%%%%%%%%%%%%%%%%%%%%%%%%%%%%%%%%%%%%%%%%%%%%%%%%%%%%%%%%%%%%%%%%%%%%%%%%%%%%%%%%%%%%%%%%%%%%%%%%%%%%%%%%%%%%%%%%%%%%%%%
\subsubsection{T-shaped H$_2\cdots$ H$_2$}
%%%%%%%%%%%%%%%%%%%%%%%%%%%%%%%%%%%%%%%%%%%%%%%%%%%%%%%%%%%%%%%%%%%%%%%%%%%%%%%%%%%%%%%%%%%%%%%%%%%%%%%%%%%%%%%%%%%%%%%%%%
In the H$_2\cdots$ H$_2$ model system, one of the H$_2$ molecules is stretched beyond its equilibrium distance and hence requires a multi-determinant description of its ground state. 
Simultaneously, the interaction energy between the subsystems is very sensitive to the choice of the dynamic electron correlation correction. 
These two features make this system an ideal candidate for testing new quantum chemistry methods designed to model static and dynamic correlation.
Furthermore, due to its small size, FCI reference calculations are computationally feasible and can be used to assess the accuracy and reliability of any approximate electron correlation method. 

This particular model system was very recently studied by Pastorczak {\em et al.}~\cite{pastorczak2017} and Hapka {\em et al.}~\cite{hapka-jctc-2019}.
Specifically, they considered the T-shaped ($C_{2v}$ point group symmetry) H$_2\cdots$H$_2$ system in which the hydrogen molecule positioned on the $C_2$ axis was being stretched.
This model system was investigated with GVB-based methods and augmented with an in-depth SAPT analysis~\cite{korona2008,korona2009,korona2008second}. 
The authors concluded that the H$_2\cdots$H$_2$ system is bound by dispersion forces and represents a stringent test case to assess the coupling between dynamic and static correlation, whose importance increases in the dissociation limit of the stretched hydrogen molecule.

\begin{figure}[tb]
	\includegraphics[width=0.5\textwidth]{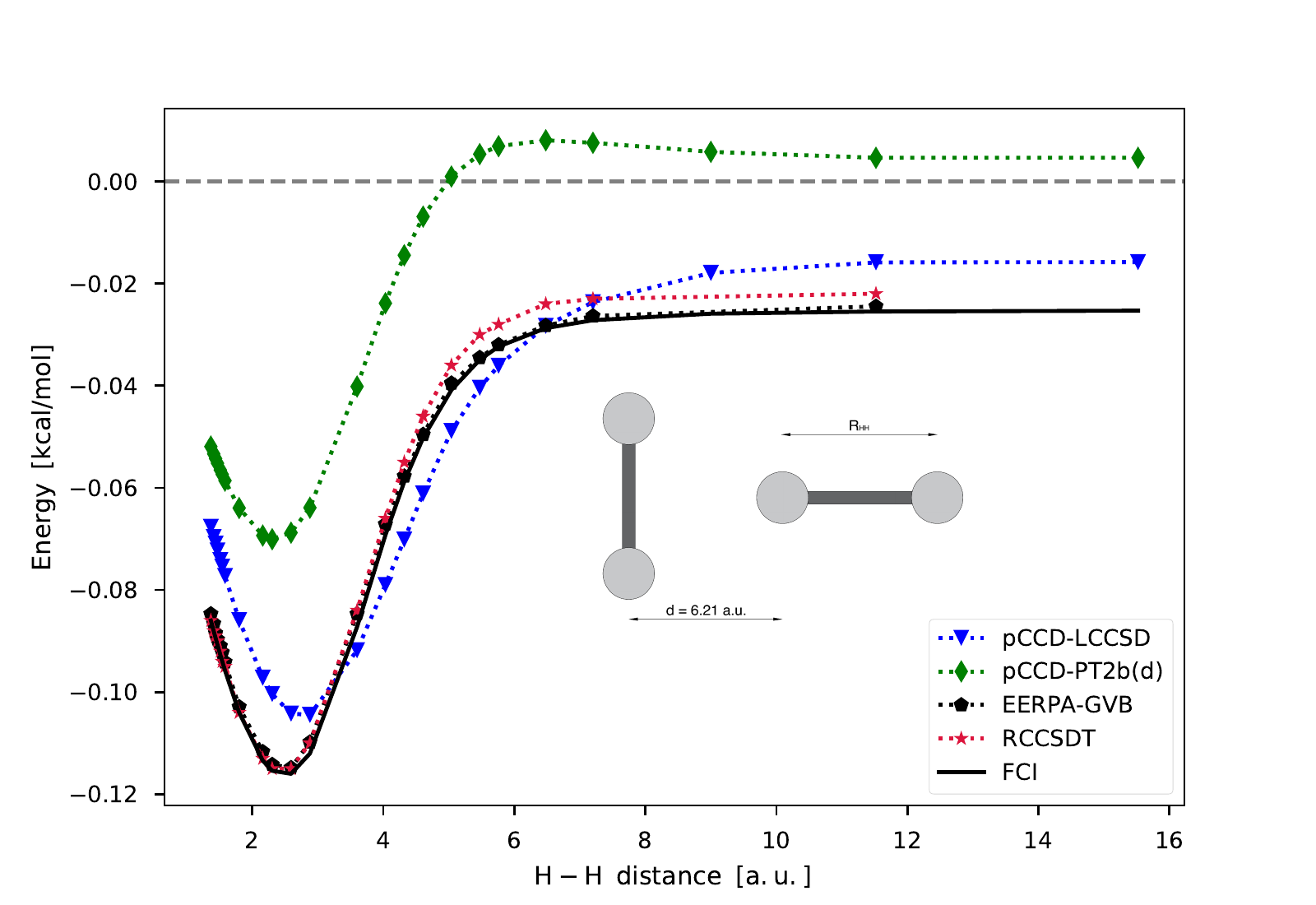}
	\caption{ Interaction energy of T-shaped H$_2\cdots$ H$_2$. Reference data for EERPA-GVB and RCCSDT is taken from Ref.~\citenum{pastorczak2017}.}
	\label{fig:H2H2}
\end{figure}

Our results for post-pCCD approaches are presented in Figure~\ref{fig:H2H2}. They are further compared to EERPA-GVB and RCCSDT reference data~\cite{pastorczak2017}. 
pCCD-LCCSD captures about 90\% of the total interaction energy around the equilibrium geometry (absolute difference of 0.012 kcal/mol) and in the dissociation limit (absolute difference of 0.01 kcal/mol at $R=15.0$ a.u.) with respect to FCI. 
In the intermediate region, however, the pCCD-LCCSD interaction energy overestimates the FCI interaction energy by about 0.01 kcal/mol. 
This overcorrelation results from the fact that pCCD (and hence pCCD-LCCSD) is exact in the numerical regime of monomers built from two electron systems, while it is not exact for the dimer containing four electrons (see Table S5 of the Supporting Information) and in the case of the dimer it gives lower total energies than FCI.
What is missing in the pCCD-LCCSD method are triple excitations (related to connected and some disconnected clusters contributions). 
As we already mentioned in previous section,  the contribution of triple excitations to the dispersion energy is always negative, 
thus, the interaction energy calculated according to Eq.~\eqref{eq:eint_supermolecular} is underestimated for the equilibrium distances of H$_2$ molecules due to the fact that the dispersion 
interaction is dominant in this system.
We should also note that pCCD-LCCSD is not variational (similar to other CC-based methods) and might predict more than 100\% of the correlation energy, hence 
it possible that the interaction energy is overestimated in the  intermediate regime.   
Furthermore, in this intermediate regime, dispersion is less dominant and thus the missing triple excitations are compensated by intramonomer contributions to electrostatic interactions. This may lead to the observed overestimation in interaction energies (see also Ref.~\mbox{\citenum{pastorczak2017}}).

The results obtained from the perturbative PT2b(d) correction are rather poor for the T-shaped H$_2\cdots$H$_2$ model system compared to all other investigated methods (see Figure~\ref{fig:H2H2}).
For all considered geometries, pCCD-PT2b(d) consistently underestimates the interaction energy by about 0.4 kcal/mol.
The largest differences are observed around the equilibrium geometry and amount to 40\%. 
Moreover, the interaction energy limit is not correctly reproduced by the pCCD-PT2b(d) approach as the H$_2$+H interaction energy appears to be repulsive, as opposed to more accurate theories.
Nonetheless, pCCD-PT2b(d) can qualitatively reproduce the overall shape of the interaction well up to an H$\cdots$H distance of approximately 5 a.u..
Finally, the observed non-parallelity error (NPE) for pCCD-PT2b(d) is 0.020 kcal/mol and slightly worse than the corresponding NPE of pCCD-LCCSD (0.018 kcal/mol).
The slightly larger NPE predicted by pCCD-PT2b(d) can be associated with the artificial bump located at a distance of 5--7 a.u.

%%%%%%%%%%%%%%%%%%%%%%%%%%%%%%%%%%%%%%%%%%%%%%%%%%%%%%%%%%%%%%%%%%%%%%%%%%%%%%%%%%%%%%%%%%%%%%%%%%%%%%%%%%%%%%%%%%%%%%%%%%
\subsubsection{H$_2\cdots$ LiH}
%%%%%%%%%%%%%%%%%%%%%%%%%%%%%%%%%%%%%%%%%%%%%%%%%%%%%%%%%%%%%%%%%%%%%%%%%%%%%%%%%%%%%%%%%%%%%%%%%%%%%%%%%%%%%%%%%%%%%%%%%%
The H$_2\cdots$LiH complex with the stretched LiH molecule represents another prototype system for testing the interplay between static and dynamic correlation. 
The interaction energy of this model system strongly depends on the proper dissociation limit of the LiH unit, but is also altered by the induction and electrostatic energy contributions. 
Any deviation from the neutral separation of the LiH molecule into its individual products will generate a nonphysically high value of the dipole moment, which strongly polarizes H$_2$. 
As a result the description of the interaction energy might be completely wrong.
In contrast to \mbox{H$_2 \cdots$H$_2$}, the dispersion interaction is weaker than electrostatics and induction.
Thus, the overall performance of the tested pCCD and post-pCCD methods should be better than in the molecular hydrogen dimer (in terms of relative errors).

In our studies we adopted a model T-shaped structure, where the Li atom is facing the center of mass of the H$_2$ molecule at a distance of 4.0 a.u. 
The LiH bond is then stretched until its dissociation limit. 
Simultaneously, the H$_2$ molecule remains fixed in its equilibrium structure for every single point considered along the potential energy surface (see the Supporting Information for more details). 
%A schematic representation of this dissociation process is depicted in Figure~\ref{fig:XX}. \pawel{Make that Figure.} 
Within a single-determinant description (as in Hartree--Fock theory) of the stretched LiH molecule, the corresponding dissociation limit would conform to Li$^+\cdots$H$^-$ and thus result in a very large dipole moment. 
Furthermore, the interaction between the H$_2\cdots$LiH subunits is dominated by a electrostatic quadrupole-dipole interaction, which decays as $R^{-4}$, where $R$ is the separation between the centers of mass of H$_2$ and LiH, respectively.
When the LiH fragment is stretched, the dipole moment predicted by HF theory linearly goes to infinity.
Thus, the electrostatic interaction for $E_{\rm HF}^{\rm int}$ decays as $R^{-3}$. 
For an exact theory, the LiH molecule dissociates into two neutral atoms.
Hence, we expect that the interaction energy very quickly flattens out as soon as the Li and H atoms stop overlapping, while the interaction energy goes to the Li+H$_2$ limit and becomes repulsive.

\begin{figure}[tb]
	\includegraphics[width=0.5\textwidth]{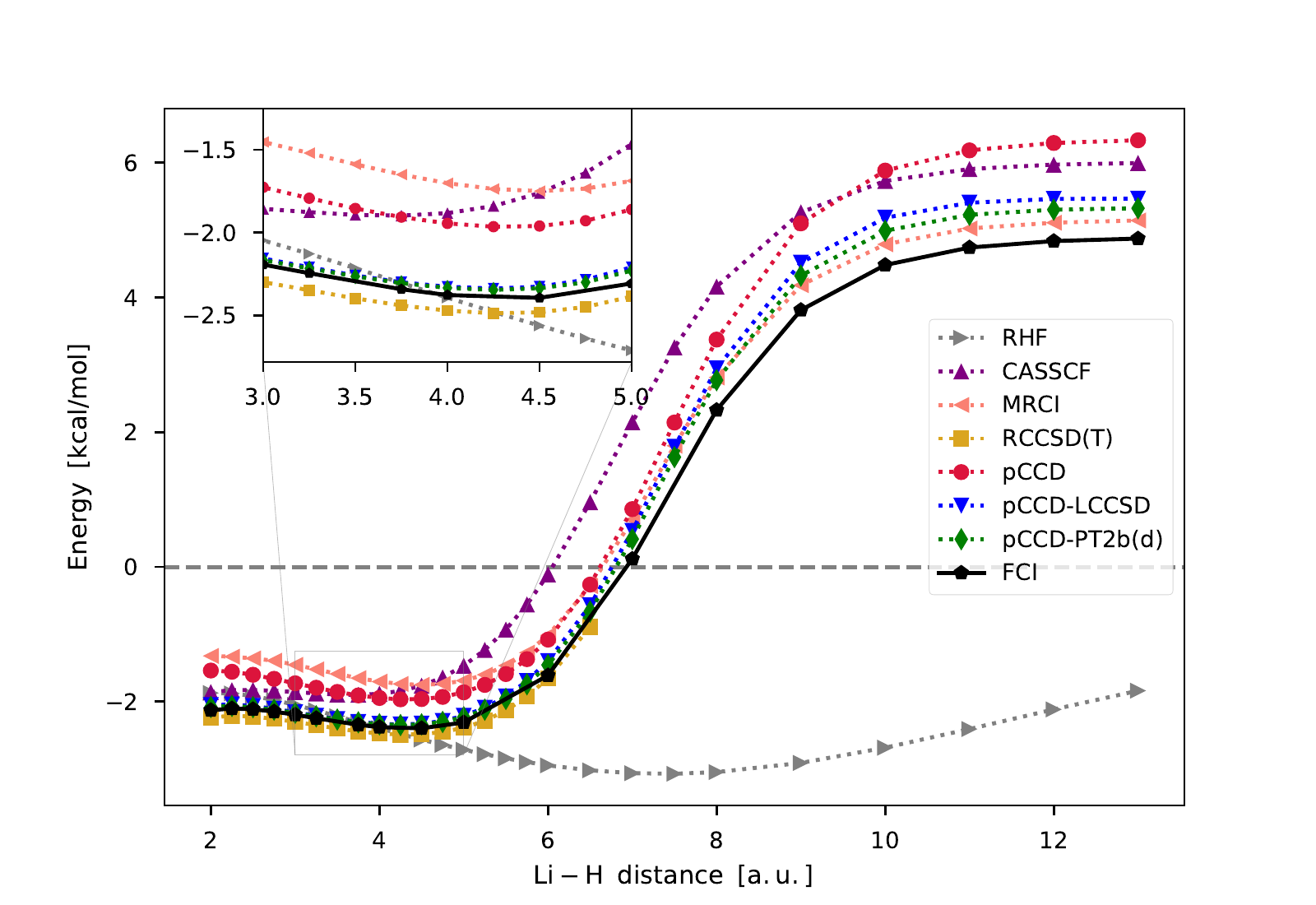}
	\caption{ Interaction energy of H$_2\cdots$ LiH. }
	\label{fig:LiHH2}
\end{figure}

Numerical results for the H$_2\cdots$LiH complex are shown in Figure~\ref{fig:LiHH2}.  
We compare CC-based interaction energies to MRCI-SD and CASSCF calculations as well as to FCI results. 
As predicted above, this model system has only little contributions of dispersion interaction given the rather small differences between pCCD/CASSCF and other methods that include dynamic correlation.
As expected, RHF features a wrong asymptotic behaviour of the interaction energy, which clearly does not flatten out.
If the active space is properly chosen, CASSCF yields an interaction energy that goes to the correct limit, while for large interatomic Li$\cdots$H separations (>10 a.u.) it becomes almost flat. 
Furthermore, the pCCD interaction energy is very close to CASSCF results: it is slightly lower in the intermediate region (5-10 a.u.), while the limit lies slightly above the limit predicted by CASSCF.
As expected, the pCCD-LCCSD interaction energy curve lies slightly above RCCSD(T) results (absolute error of 0.15 kcal/mol) in the region of small  Li--H separations (4 a.u.), but is closer to the FCI curve (absolute error of 0.6 kcal/mol) in the dissociation limit (13 a.u.) than the CASSCF interaction energy (absolute error of 1.1 kcal.mol).
MRCI-SD calculations improve the CASSCF description of the interaction energy and reduce the absolute error to 0.27 kcal/mol with respect to FCI.
Surprisingly, pCCD-PT2b(d) again outperforms pCCD-LCCSD in both regions by 0.01 and 0.15 kcal/mol for small Li--H separations and in the vicinity of dissociation, respectively.
This also translates into smaller NPEs for pCCD-PT2b(d), which yields an NPE of 0.482 kcal/mol compared to 0.667 kcal/mol predicted by pCCD-LCCSD.

%The purpose of studying this system is to show that  the methods studied here properly describe the avoided crossing.
%It is known that H$_3^+\cdots$ H$_2$ 
%%%%%%%%%%%%%%%%%%%%%%%%%%%%%%%%%%%%%%%%%%%%%%%%%%%%%%%%%%%%%%%%%%%%%%%%%%%%%%%%%%%%%%%%%%%%%%%%%%%%%%%%%%%%%%%%%%%%%%%%%%%
\subsubsection{H$_3^+\cdots$ H$_2$}
\begin{figure}[tb]
	\includegraphics[width=.4\textwidth]{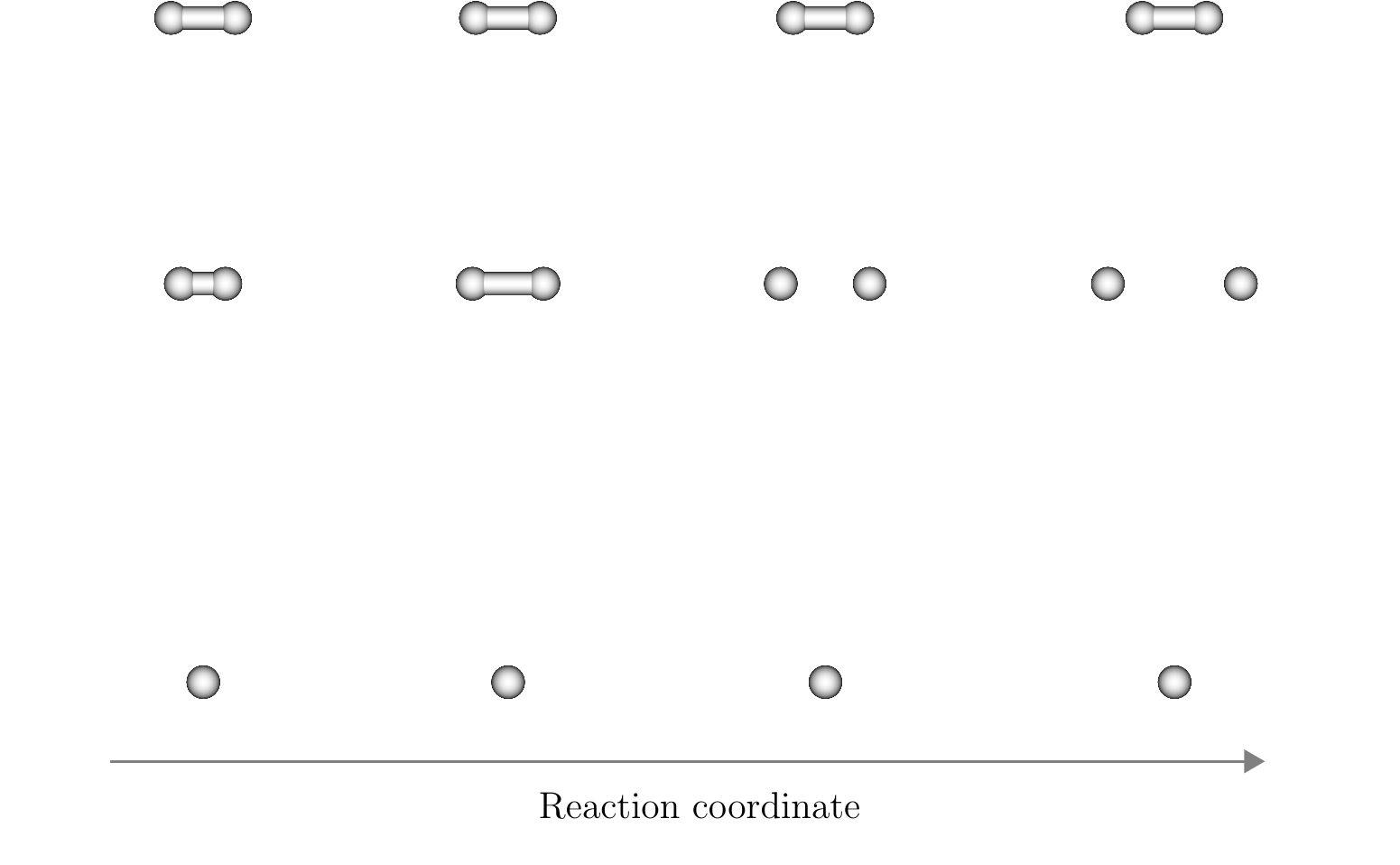}
	\caption{Schematic dissociation process of H$_3^+\cdots$ H$_2$.}
	\label{fig:pesh2h3}
\end{figure}

%%%%%%%%%%%%%%%%%%%%%%%%%%%%%%%%%%%%%%%%%%%%%%%%%%%%%%%%%%%%%%%%%%%%%%%%%%%%%%%%%%%%%%%%%%%%%%%%%%%%%%%%%%%%%%%%%%%%%%%%%%%
For our final test system, we have chosen the symmetric stretching of the \ce{H2} moiety in the { H$_3^+\cdots$ H$_2$ } supermolecule.
Specifically, the molecular geometry of the H$_3^+$ unit can be arranged in such a way that the supermolecular system passes through an avoided crossing when the \ce{H2} bond (of the H$_3^+$ unit) is stretched (see Figure~\ref{fig:pesh2h3} for more details concerning the reaction pathway).
That is, we can pick a molecular geometry for the H$_3^+$ monomer near which the supermolecular system exhibits an avoided crossing~\cite{aguado2010,aguado2000,prosmiti1997} between the two lowest-lying configurations.
These configurations correspond to the (H$^+\cdots$H$_2$) and (H$\cdots$H$_2^+$) dissociation channels. 
Due to the avoided crossing, the H$_3^+\cdots$ H$_2$ dissociation process features a sudden change in the chemical character and electronic wavefunction along the chosen reaction coordinate.
In this specific setup (see also Figure~\ref{fig:pesh2h3}), the interaction between the individual subsystems changes from (H$^+\cdots$H$_2$ and H$_2$) to (H$\cdots$ H$_2^+$ and H$_2$).
Thus, two interaction regimes are present: in the former case, dispersion forces dominate, while in the latter case induction takes over.
Here, we will explore the crossover between those two regimes more closely.
Specifically, we will scrutinize the transition between two chemically different electronic configurations of one of the monomers.
%%%%%%%%%%%%%%%%%%%%%%%%%%%%%%%%%%%%%%%%%%%%%%%%%%%%%%%%%%%%%%%%%%%%%%%%%%%%%%%%%%%%%%%%%%%%%%%%%%%%%%%%%%%%%%%%%%%%%%%%%%%
%%%%%%%%%%%%%%%%%%%%%%%%%%%%%%%%%%%%%%%%%%%%%%%%%%%%%%%%%%%%%%%%%%%%%%%%%%%%%%%%%%%%%%%%%%%%%%%%%%%%%%%%%%%%%%%%%%%%%%%%%%%

The reaction coordinate of the H$_3^+\cdots$ H$_2$ model system is constructed in a specific way: the H$^+_3$ unit is strongly stretched to form an isosceles triangle such that the distant H atom is separated from the (center of mass of the) remaining H atoms by 9 a.u.
The interatomic distance in the \ce{H2} fragment of the \ce{H3+} unit is then being varied, while the H$_2$ monomer is kept fixed in its equilibrium position throughout the reaction pathway (see also Figure~\ref{fig:pesh2h3}) and is positioned at 6 a.u. with respect to the center of mass of the individual \ce{H2} fragments.
For interatomic distances smaller than 2.4 a.u., the H$^+_3$ unit features a neutral H$_2$ molecule that is polarized by a proton, while for stretched H--H bond lengths a charge transfer occurs and the H$^+_3$ system is composed of a neutral H atom that interacts with a H$_2^+$ molecular ion.
In the former case, the dominant interaction energy of the H$^+_3$ unit with the H$_2$ molecule is the dispersion energy between the H$_2$ fragments, while in the latter case the induction energy significantly increases and becomes the dominant interaction contribution as the positive charge on the H$_2^+$ ion strongly polarizes the H$_2$ molecule. 

\begin{figure}[tb]
	\includegraphics[width=0.5\textwidth]{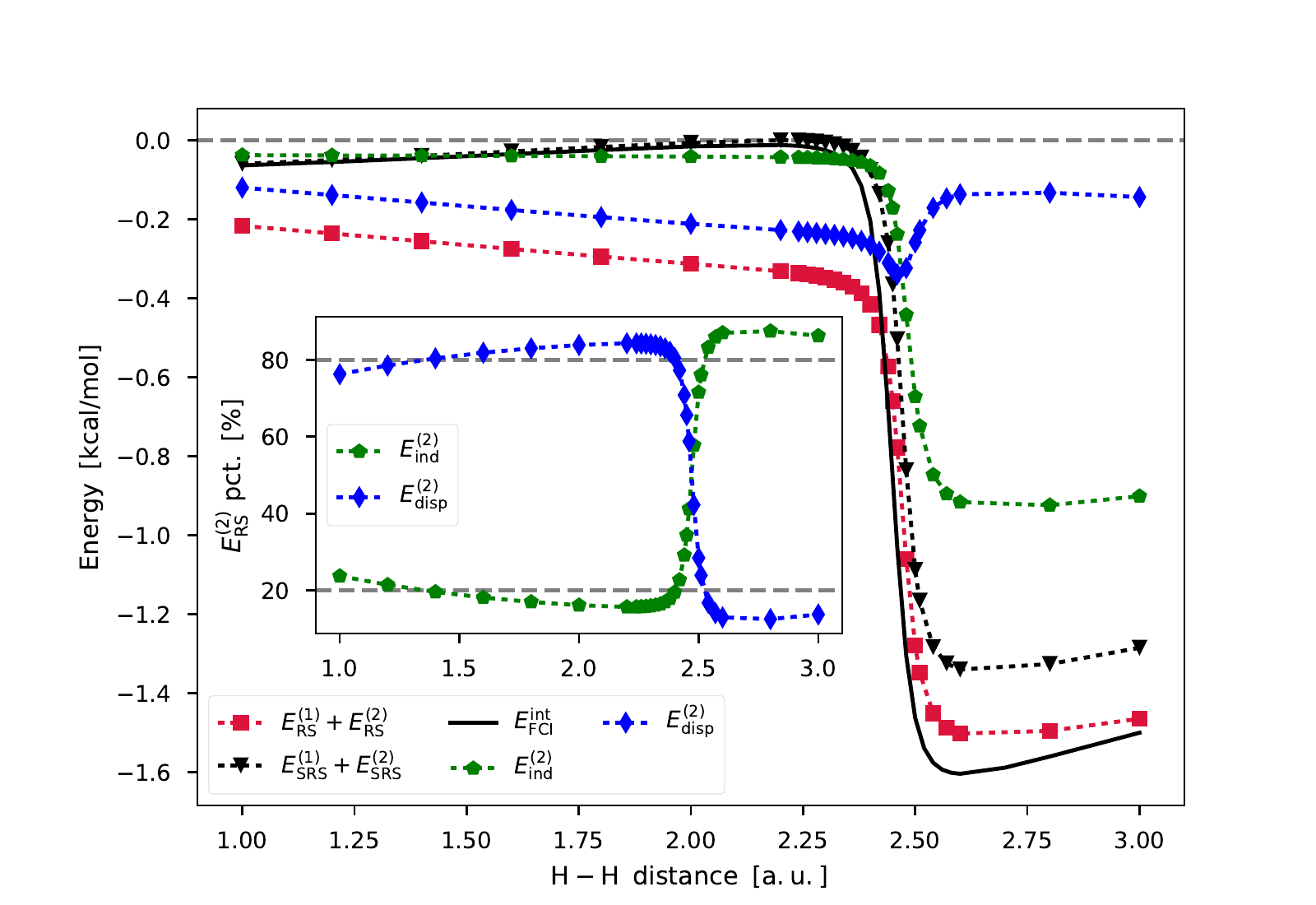}
	\caption{ Second order interaction energy of H$_3^+\cdots$ H$_2$. A basis set of aug-cc-pVDZ quality was employed due to software limitations. Inset: percentage contributions of $E^{(2)}_{\rm disp}$ and $E^{(2)}_{\rm ind}$ to the second order interaction energy -- $E^{(2)}_{\rm RS}$. }
	\label{fig:H3+H2_sapt2}
\end{figure}

To scrutinize the transition between the individual contributions to the interaction energy, we have performed second order SAPT (RS and Symmetrized Rayleigh--Schr{\"o}dinger (SRS)) calculations based on FCI 2-electron wavefunctions of the H${^+_3}$ and H${_2}$ monomers.
Finite-field FCI calculations of the respective monomers allow us then to obtain the second order induction.
Knowing that the second-order RS energy can be decomposed into dispersion and induction, we are able to retrieve second-order dispersion as the difference between the RS and induction energy.
Moreover, the SRS counterpart will provide some insights into second order exchange effects.
The corresponding numerical results are presented in Figure~\ref{fig:H3+H2_sapt2}.
For distances smaller than 2.4 a.u., $E^{(1)}_{\rm SRS} + E^{(2)}_{\rm SRS}$ perfectly agrees with the interaction energy obtained from supermolecular FCI.
Nevertheless, dispersion dominates $E^{(2)}_{\rm RS}$ in this region contributing 75-85\% of the total interaction energy (see inset of Figure~\ref{fig:H3+H2_sapt2}), whereas induction is responsible for the remaining 15-25\%.
After the crossing point ($r_{\textrm{H-H}} > 2.50$ a.u.), the situation changes completely.
Corrections calculated from up to second-order SRS do no longer match with FCI results, highlighting the importance of higher-order contributions (e.g., third-order induction). Note the small kink on the plot of the dispersion energy, which is due to the small energy denominator close to the avoided crossing of \mbox{\ce{H3+}}. Having understood the nature of the interaction and the progression of the interaction energy in the H$_3^+\cdots$ H$_2$ model system, we can now assess the quality of pCCD-based approaches. 

\begin{figure}[tb]
	\includegraphics[width=0.5\textwidth]{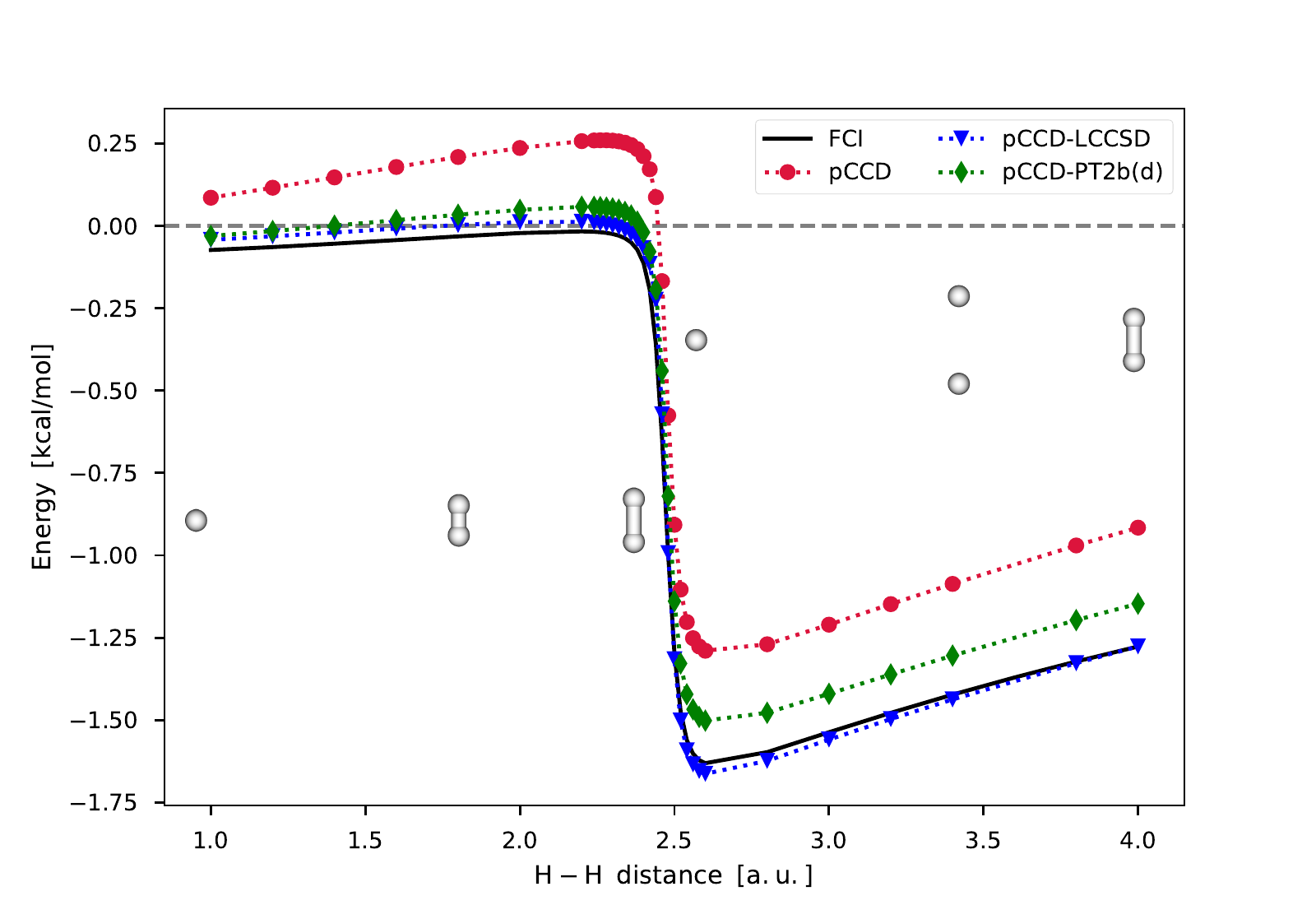}
	\caption{ Supermolecular interaction energy of H$_3^+\cdots$ H$_2$. Note that except for very short \mbox{\ce{H-H}} distances, we were not able to converge RHF calculations and hence RHF results are not shown in the Figure.}
	\label{fig:H3+H2_pes}
\end{figure}

Figure~\ref{fig:H3+H2_pes} shows the interaction energy obtained from pCCD, pCCD-LCCSD, and pCCD-PT2b(d) compared to the FCI reference.
As one would expect, pCCD follows the shape of the FCI curve.
However, due to the lack of a proper description of dynamic correlation, it severely underestimates the attractive aspect of the interaction (difference up to 0.3 kcal/mol).
Adding dynamic correlation using \textit{a posteriori} corrections greatly improves the performance.
In the dispersion-dominated region, differences with respect to the FCI results are reduced by one order of magnitude if dynamic correlation is included on top of the pCCD wavefunction (errors amount to 0.03 kcal/mol).
For the induction-dominated part of the reaction coordinate, the LCCSD correction slightly overcorrelates (see Table S9 of the Supporting Information) the total energy of the dimer resulting in a stronger attraction in pCCD-LCCSD compared to FCI results for interatomic separations of $2.5$ to $3.5$ a.u.
In contrast to the H$_2\cdots$LiH model system, PT2b(d) does not outperform LCCSD, yielding difference of about 0.06 kcal/mol in the dispersion and about 0.12 kcal/mol in induction branch, respectively.

%\begin{figure}[tb]
%	\includegraphics[width=.8\textwidth]{./figs/streched/h3+h2_sapt2_contributions.pdf}
%	\caption{ Decomposition of second order interaction energy of H$_3^+\cdots$ H$_2$ into dispersion and induction contributions }
%	\label{fig:H3+H2_sapt2_contributions}
%\end{figure}

%Quantum chemical studies by Karpfen~\cite{Karpfen_1999,Karpfen_2000,Karpfen-jpca} indicates high sensitivity of the equilibrium structure and properties of NH$_3$--F$_2$ to the choice of basis set and electron correlation method. 
%
%\begin{figure}\label{fig:NH3F2}
%	\includegraphics[width=1.0\textwidth]{./figs/streched/nh3_f2_pccd_flavours.png}
%	\caption{Interaction energy of H$_2\cdots$ H$_2$. \filip{Just for disscusion purpose will be overhauled}}
%\end{figure}
%%%%%%%%%%%%%%%%%%%%%%%%%%%%%%%%%%%%%%%%%%%%%%%%%%%%%%%%%%%%%%%%%%%%%%%%%%%%%%%%%%%%%%%%%%%%%%%%%%%%%%%%%%%%%%%%%%%%%%%%%%
\section{Conclusions and Outlook}\label{sec:conclusions}                       
%%%%%%%%%%%%%%%%%%%%%%%%%%%%%%%%%%%%%%%%%%%%%%%%%%%%%%%%%%%%%%%%%%%%%%%%%%%%%%%%%%%%%%%%%%%%%%%%%%%%%%%%%%%%%%%%%%%%%%%%%%
                                                                                
In this article, we have investigated the performance of some recently developed post-pCCD methods in describing interaction energies in non-covalent compounds and non-covalent complexes that feature covalent bond breaking.
To reliably model chemical reactions, an (approximate) quantum chemistry method has to account for both types of electron correlation effects: static and dynamic. 
Only in this case, it is possible to correctly reproduce equilibrium bond lengths, dissociation limits, and possible bond-breaking/forming scenarios.
Given that most of the reactive potentials used in molecular dynamics simulations are obtained from expensive and complicated MRCI or perturbation theory-based approaches, we seek new types of methods that are robust, inexpensive, and at the same time rigorously size-consistent.
                                                                                
In this work, we first scrutinized the performance of pCCD and post-pCCD-based methods in modeling dispersion- and hydrogen-bonded systems present in the A24 set.
We showed that the pCCD interaction energies do not include any dispersion energy and that pCCD gives similar results as dispersion-free SAPT (up to the second order, augmented by the $\delta_{\rm HF}^{(2)}$ term). The largest differences between these two approaches amount to 0.55 kcal/mol. 
The best performance of post-pCCD approaches was observed for pCCD-PT2b(d) and pCCD-LCCSD.
Both methods correctly model the dispersion interaction effects. 
The absence of triply-excited contributions in the LCCSD correction (related to connected and some disconnected clusters contributions) leads to a systematic underestimation of the interaction energy (with a ME of $-$0.284 kcal/mol), which is similar to CCSD (with an error of about $-$0.125 kcal/mol) and CEPA(0) (with a ME of $-$0.125 kcal/mol) interaction energies.
Surprisingly, the statistically best performance for the A24 data set was obtained with the PT2b(d) approach, resulting in the smallest ME of approximately $-$0.118 kcal/mol.
The remaining PT2b models yield similar, albeit slightly worse results. 
All the other perturbative corrections on top of pCCD (PT2SDd, PT2MDd, PT2SDo, PT2MDo) give errors in the order of 1 kcal/mol or more (See Table S1 of the Supporting Information).
                                                                                
In the second part of our study, we focused on modeling interaction energies in complex model systems featuring bond-breaking processes: the T-shaped H$_2$ dimer, the LiH$\cdots$H$_2$ system, and the H$_3^+\cdots$H$_2$ complex.
The bare pCCD interaction energy in the T-shaped H$_2$ dimer is purely repulsive. 
The more sophisticated pCCD-LCCSD model properly reproduces the shape of the potential energy. 
However, the pCCD-LCCSD interaction energy is underestimated by about 0.1 kcal/mol around the minimum and in the vicinity of dissociation. 
In the intermediate region, it slightly overestimates the reference FCI results. Nonetheless, the overall differences are very small (< 0.1 kcal/mol).
The less expensive pCCD-PT2b(d) approach properly reproduces the shape of the potential of the T-shaped H$_2$ dimer around its equilibrium structure.
Closer to the dissociation limit, the pCCD-PT2b(d) potential, however, deviates from the reference potential more significantly. 
Furthermore, the complete pCCD-PT2b(d) potential energy surface is shifted by about 0.4 kcal/mol compared to the FCI reference curve.
Given the performance of post-pCCD methods on the A24 set, these result are understandable. 
We believe that inclusion of triple excitations in these models would improve the description of dispersion interactions and the overall performance.

The T-shaped LiH$\cdots$H$_2$ complex is dominated rather by induction interaction than dispersion forces, as opposed to the H$_2$ dimer. 
Hence, the overall better agreement of pCCD-LCCSD with FCI for this system was to be expected. 
The errors in interaction energies, however, increase for larger Li--H bond distances (the error changes from 0.02 to 0.6 kcal/mol). 
pCCD-PT2b(d) yields similar errors with respect to FCI as pCCD-LCCSD, albeit its performance is slightly better throughout all distances (with errors up to 0.15 kcal/mol).

Finally, we assessed the performance of various post-pCCD methods in describing the avoided crossing in the H$_3^+\cdots$H$_2$ system that features a slightly distorted H$_3^+$ ion molecular structure.
The contributions from dispersion and induction rapidly change as the \ce{H2} fragment in the H$_{3}^{+}$ unit is stretched and correspond to either H$^+\cdots$H$_2$ (dispersion branch) or H$\cdots$H$_2^+$ (induction branch).
The pCCD method systematically underestimates the interaction energy by about 0.2 kcal/mol and 0.35 kcal/mol in the dispersion and induction branch, respectively.
For the dispersion branch, pCCD-LCCSD slightly underestimates the FCI interaction energy, which can be explained by neglecting triple excitation in the cluster operator.
The opposite is true for the induction branch, where pCCD-LCCSD slightly overestimate the interaction energy.
The PT2b(d) correction improves the pCCD interaction energies in both branches, but its performance in the induction branch is slightly worse compared to the LCCSD correction.

To conclude, our study highlights that the pCCD-LCCSD method is very competitive in terms of accuracy compared to other methods that are currently used to model bond-breaking processes, such as MRCI-SD or CASPT2.
The method is size-consistent and quite robust: we do not need to define active spaces and a set of linear equations is solved for the cluster amplitudes.
Furthermore, such pCCD-based methods (linearized or not) can be systematically improved by considering higher-order excitation operators in the cluster operator, like triples, etc.
Since pCCD is dispersion-free, the development of a fully-fledged SAPT methodology based on a pCCD description of the monomers is promising.
Some initial work in this particular direction has been recently presented by Hapka \textit{et al.}~\cite{hapka-jctc-2019}, who show that extended RPA might be a suitable framework for the derivation of second-order contributions.
%%%%%%%%%%%%%%%%%%%%%%%%%%%%%%%%%%%%%%%%%%%%%%%%%%%%%%%%%%%%%%%%%%%%%%%%%%%%%%%%%%%%%%%%%%%%%%%%%%%%%%%%%%%%%%%%%%%%%%%%%%
\section*{Supporting Information}
%%%%%%%%%%%%%%%%%%%%%%%%%%%%%%%%%%%%%%%%%%%%%%%%%%%%%%%%%%%%%%%%%%%%%%%%%%%%%%%%%%%%%%%%%%%%%%%%%%%%%%%%%%%%%%%%%%%%%%%%%%
The Supporting Information is available free of charge on the ACS Publications website at DOI:.

Coordinates and energies for the dissociation process of model systems with mixed dynamic and static correlation energy (H$_2\cdots$ H$_2$, H$_2\cdots$ LiH, and H$_3^+\cdots$ H$_2$). 
A modified aug-cc-pVTZ basis set for H. 
A24 statistical errors for all variants of perturbation theory corrections on top of pCCD.  
%%%%%%%%%%%%%%%%%%%%%%%%%%%%%%%%%%%%%%%%%%%%%%%%%%%%%%%%%%%%%%%%%%%%%%%%%%%%%%%%%%%%%%%%%%%%%%%%%%%%%%%%%%%%%%%%%%%%%%%%%%
\section*{Acknowledgement}
%%%%%%%%%%%%%%%%%%%%%%%%%%%%%%%%%%%%%%%%%%%%%%%%%%%%%%%%%%%%%%%%%%%%%%%%%%%%%%%%%%%%%%%%%%%%%%%%%%%%%%%%%%%%%%%%%%%%%%%%%%
F.B.,~P.T.,~and~P.Sz.\.{Z}~thank an OPUS grant of the National Science Centre, Poland, (no.~2015/19/B/ST4/02707). 
P.T.~thanks a POLONEZ 1 grant financed by Marie-Sk\l{}odowska-Curie COFUND. This project has received funding from the European Union's Horizon 2020 research and innovation programme under the Marie Skłodowska-Curie grant agreement No 665778. 
K.B.~acknowledges financial support from a Marie-Sk\l{}odowska-Curie Individual Fellowship project no.~702635--PCCDX and a scholarship for outstanding young scientists from the Ministry of Science and Higher Education.

Calculations have been carried out using resources provided by Wroclaw Centre for Networking and Supercomputing (http://wcss.pl), grant No.~411.

%\bibliography{rsc}
\end{document}